\documentclass{iopjournal}

\usepackage[T1]{fontenc}
\usepackage{newunicodechar}
\newunicodechar{≠}{\neq}
\newunicodechar{⁓}{\texttildelow{}}

\usepackage[a4paper,margin=1in,includehead]{geometry}
\usepackage{svg}
\usepackage{textgreek}
\usepackage{chemformula}
\usepackage{multirow}
\usepackage{booktabs}
\usepackage{threeparttable}
\usepackage{tabularx}
\usepackage{float}
\usepackage{subcaption}
\usepackage{makecell}
\usepackage{array}
\usepackage{siunitx}
\usepackage{cite}

\makeatletter

\providecommand\received[1]{}
\providecommand\revised[1]{}
\providecommand\accepted[1]{}
\providecommand\published[1]{}
\makeatother

\makeatletter
\long\def\@addmarginpar#1{} 
\long\def\@xympar#1{}       
\renewcommand\marginpar[1]{}
\makeatother

\begin{document}

\articletype{Paper} 

\title{Bernal Stacking and Symmetry-Inequivalent Antiferromagnetism in MSi$_2$N$_4$ Heterobilayers}

\author{Brandon Pedroza-Rojas$^{1,2,*}$\orcid{0000-0002-2168-7388}, David W. Facemyer$^{1,*}$\orcid{0000-0002-5146-0299} and Ariadna Sánchez-Castillo$^{3}$\orcid{0000-0001-6564-9837}}

\affil{$^{1}$Department of Physics and Astronomy and Nanoscale and Quantum Phenomena Institute, Ohio University, Athens, Ohio 45701, USA}

\affil{$^{2}$Institute of Basic Sciences and Engineering, Autonomous University of the State of Hidalgo, Mineral de la Reforma, Hidalgo 42184, Mexico}

\affil{$^{3}$Superior School of Apan, Autonomous University of the State of Hidalgo, Chimalpa Tlalayote, Hidalgo 43920, Mexico}

\affil{* These authors contributed equally to this work.}

\email{bp297224@ohio.edu, pe484939@uaeh.edu.mx}

\email{df008219@ohio.edu}

\email{ariadna\_sanchez@uaeh.edu.mx}


\begin{abstract}
Layered MA$_2$Z$_4$ compounds, structural relatives of MoS$_2$ discovered in 2020, exhibit rich magnetic behavior arising from reduced dimensionality, noncentrosymmetric lattice symmetries, and stacking-dependent exchange interactions. Here, we investigate Bernal-like stackings in H-phase MA$_2$Z$_4$ (M = Mn and Fe; A = Si; Z = N) monolayers and bilayers by combining first-principles spin-dependent relaxation energies with a localized-spin Heisenberg description. From density-functional calculations, we extract the dominant intralayer exchange couplings up to third-nearest neighbors and the leading interlayer exchanges up to second-nearest neighbors, enabling construction of an effective bilayer spin Hamiltonian. We first analyze interface-driven proximity effects within a ferromagnetic reference configuration, demonstrating how recovery of AB-type stacking and spin alignment--while varying only the transition-metal species--provides a route for selectively tuning magnetic order and symmetry breaking within the P$\bar{6}$m2 space group. Building on this microscopic understanding of the bonding environment, we then examine antiferromagnetic ordering tendencies in the coupled layers. Exact diagonalization of the resulting bilayer Hamiltonian reveals the magnetic ground state and low-lying excitation spectrum, showing that the interlayer exchange is not merely perturbative but competes directly with intralayer interactions in stabilizing the observed spin configurations. These results establish Bernal-stacked MA$_2$Z$_4$ bilayers as a platform in which stacking geometry and exchange hierarchy jointly govern magnetic reconstruction, offering a controlled pathway toward domain selection and spin-texture engineering in low-dimensional van der Waals materials.
\end{abstract}

\section{Introduction}
Van der Waals heterostructures (vdWHs) serve as essential frameworks for vertical-engineering in spintronic devices \cite{liu2021,jia_spintronic_2025}, integrating magneto-resistive junctions to optimize quantum phenomena \cite{39xc-2f3m}. Recent advances in spin-orbit torque and magnonics have demonstrated room-temperature antiferromagnetic (AFM) tunneling junctions that circumvent ferromagnetic limitations, such as incomplete polarization and field sensitivity\cite{zhong2024integrating,PhysRevB.110.L220409,Sun2024Dipolar,Qin2023RoomTemperature}. These enable field-driven polarization with tunable magnetoresistance across antiparallel, parallel, and spin-inclination transitions\cite{Chen2023OctupoleDriven,PhysRevApplied.21.054002}, complementing AFM spintronic paradigms\cite{Nemec2018AntiferroOptospintronics,Sierra2021vdWSpintronics}. AFM insulators further harness thermal gradients for spin-current generation via the spin Seebeck effect\cite{PhysRevLett.116.097204}.

Current demands for faster, durable electronics with sub-particle control\cite{Dey2021Spintronics} drive exploration of AFM semiconductors hosting coherent magnon-exciton interactions\cite{Wang2021SpinPL,Kang2020NiPS3}, magnon-driven skyrmions\cite{PhysRevB.104.054419}, interfacial AFM skyrmions\cite{He2024AFMSkyrmionNanoLett}, and magnon-phonon coupling\cite{Telford2020CrSBr,Lee2021CrSBr,Wilson2021Interlayer}. These phenomena emerge from collinear (Néel/stripe) or non-collinear orders\cite{Olsen2024Antiferromagnetism2D}, stabilized by spin-orbit coupling (SOC)\cite{https://doi.org/10.1002/adma.202310379}. Although graphene displays defect-enhanced magnetism\cite{PhysRevB.90.085429,Nair2013DefectMagnetismGraphene}, silicon-based spintronics exhibits suboptimal performance\cite{Sverdlov2015SiliconSpintronics}. This underperformance persists due to fundamental exchange-correlation phenomena that govern total spin magnetic moments\cite{Kohn1965SelfConsistent,Dederichs1984GroundStates,Liechtenstein1987ExchangeInteractions}.

The MA\textsubscript{2}Z\textsubscript{4} family hosts a remarkably broad range of electronic and magnetic behaviors. Schottky barriers in these compounds enable efficient quantum tunneling\cite{Wang2021OhmicContacts}, MoSi\textsubscript{2}N\textsubscript{4} supports half-metallicity and valley polarization\cite{Ai2021SpinValleyMoSi2X4}. Their vdWHs further allow giant tunnel magnetoresistance\cite{Wu2022GiantTMR} and strain-tunable Schottky responses\cite{Zhang2023Ti3C2T2TiSi2N4}. Janus variants\cite{Dey2022IntrinsicFM,Zhang2022JanusMA2Z4}, nanoribbons\cite{Zhang2022JanusMA2Z4,Liu2022StrainMoSi2N4,Shu2022PhotogalvanicDetector,Su2023MoSi2N4Nanoribbons,Zhang2024SpinFET}, and valleytronic applications\cite{Ahmad2024TiSi2N4,Yang2021ValleyPseudospin,Liang2023ValleySpinSplittings} expand this landscape, often tied to their nontrivial topological phases\cite{Wang2021MA2Z4}. Heavy-electrode configurations show high optical absorption\cite{Wang2022VSi2N4Anode}, while MnSi\textsubscript{2}N\textsubscript{4} (MSN) and FeSi\textsubscript{2}N\textsubscript{4} (FSN) exhibit metallicity in the ferromagnetic (FM) order\cite{Wu2022MoSi2N4}, pronounced optical activity\cite{Zhao2023}, and sizable strain-tunable magnetic anisotropy (MAE$=-345$ µeV/Mn)\cite{Tian2022ElectronicPropertiesMnSi2N4}.

Regardless, the stripe-type AFM breaks the in-plane rotational symmetry inherent to the hexagonal lattice\cite{Wei2025UpperCriticalFields}, promoting local distortion, end-chain defects, which enhances spin polarization\cite{Chou2024SpinPolarization} and supports time-reversal symmetry breaking\cite{Smejkal2020CrystalTRSB}. Crucially, SOC stabilizes two-dimensional magnetism despite Mermin–Wagner constraints\cite{PhysRevLett.17.1133} by inducing magnetocrystalline anisotropy\cite{PhysRevB.111.155122}. 

Using density functional theory (DFT) in the DFT-D3(BJ)$+U+J$ formulation, we characterize the monolayer-bilayer progression and identify the stripe-AFM ground state, along with its associated magnetic phase transitions. The extracted exchange landscape shows that the interlayer terms are not simply perturbative: the nearest-interlayer exchange is comparable to the dominant intralayer scale in one layer, while the next-nearest interlayer term competes directly with the weaker intralayer couplings in the other. To clarify the magnetic ordering tendencies that emerge from this hierarchy, we constructed a localized-spin model and examined its low-lying magnetic excitation spectrum through exact diagonalization. The resulting picture demonstrates how these competing exchange channels can stabilize intrinsic spin control across the two layers, enabling spin-valve-like configurations and functional FM/AFM tunnel-junction behavior in this class of bilayers (BLs).

\section{Procedural Framework}
A systematic investigation was conducted into the electronic and magnetic properties of these heterostructures. To this end, DFT+$U$ was employed, with the implementation being carried out using \textsc{Quantum ESPRESSO} 7.3.1~\cite{giannozzi2009, giannozzi2020}. The valence electron configurations for the projector augmented wave (PAW) pseudopotentials~\cite{blochl1994, joubert1999} were defined as: Mn ($3s^2 3p^6 3d^5 4s^2$; 15~$e^-$), Fe ($3s^2 3p^6 3d^6 4s^2$; 16~$e^-$), Si ($3s^2 3p^2$; 4~$e^-$), and N ($2s^2 2p^3$; 5~$e^-$).

Scalar relativistic calculations were performed within the Local Spin Density Approximation (LSDA) for structural relaxations and collinear magnetic searching. For the calculations pertaining to the magnetocrystalline anisotropy (MAE), we incorporated spin-orbit coupling (SOC). A critical aspect of this methodology is the treatment of on-site Coulomb interactions.  The Dudarev formulation~\cite{dudarev1998}, has proven to be a reliable tool for analyzing isotropic properties. However, its application is often hindered by the utilization of spherically averaged interactions, which can obscure more nuanced orbital-dependent effects. In addressing the preservation of anisotropic contributions to energy and the mitigation of suppression of orbital moments in the non-collinear regime, the rotationally invariant formulation proposed by Liechtenstein et al.~\cite{Liechtenstein1995} was employed.

Additionally, the accurate modeling of the interlayer binding in vdWHs necessitates a meticulous treatment of dispersion forces. The Grimme D3 correction was incorporated with Becke–Johnson (BJ) damping~\cite{grimme2010, grimme2011}. The BJ damping scheme is meticulously designed to effectively attenuate the dispersion correction at short ranges, thereby preventing the unphysical double-counting of correlation effects already captured by the exchange-correlation functional. Concurrently, it ensures the accurate recovery of the long-range London dispersion behavior essential for 2D stacking trends.

\section{Computational details}
Spin-polarized Density Functional Theory (known as LSDA) calculations were performed using the Quantum Espresso package (QE)~\cite{giannozzi2009,giannozzi2020}. Electron-ion interactions were described using the Projector Augmented Waves (PAW) pseudopotential method~\cite{blochl1994,joubert1999}, as in the first MA\textsubscript{2}Z\textsubscript{4}, where it was shown by the authors to best match the experimental band gap results in MoSi\textsubscript{2}N\textsubscript{4}. Additionally, exchange-correlation interactions were implemented using the Generalized Gradient Approximation (GGA) with the Perdew–Burke–Ernzerhof (PBE) parametrization~\cite{perdew1996}. 

For the treatment of van der Waals dispersions, the D3 semiempirical approach~\cite{grimme2006,grimme2010,grimme2011,grimme2016}, with Becke–Johnson damping correction~\cite{becke2005}, was employed. To describe the behavior of \( d \)-electrons, the Hubbard model was used~\cite{hubbard1963,aikebaier2015}, of the Dudarev type, with \( U_{\text{eff}} = U - J_0 \), where values of 4.6/0.6 and 6.0/0.5 eV were used for the Hubbard and Hund components in Mn/Fe, respectively, as reported~\cite{An2023}.

Electronic states were expanded using plane waves up to nearly 320 Ry and 600 Ry for the kinetic energy cutoff and the charge density cutoff, respectively, using a convergence test sequence for each monolayer, as shown in Tables~\ref{tab:table7}. A convergence criterion of \( 1 \times 10^{-8} \) Ry/Bohr was employed, with Fermi–Dirac smearing up to 0.01 Ry. To mitigate image interactions, vacuum layers of 30 and 50 Å were added in the \(z\)-direction for monolayers and bilayers, respectively.

The Brillouin zone was sampled using a Monkhorst–Pack grid~\cite{monkhorst1976}. LSDA was implemented in collinear mode and with spin–orbit coupling. Collinear calculations enabled the determination of the magnetic ground state for each magnetic monolayer and stacking configuration, as shown in Table~\ref{tab:table7}.

\begin{table*}[ht]
\centering
\caption{Initial collinear magnetization parameters: layered AFM and FM alignments in magnetic monolayers, and each stacking through minimum energy interlayer distances.}
\label{tab:table7}
\begin{tabular}{lllll}
\toprule
System & M$_{atom}$ & $E_\text{cut}^{\text{k}}$ (Ry) & $E_\text{cut}^{\rho}$ (Ry) & $k$ \\
\midrule
A1-type AFM &  &  &  &  \\
 & M$_{Mn}$ = ±1 & 120 & 480 & 5×5×1 \\
A2-type AFM &  &  & &  \\
 & M$_{Fe}$ = ±1 & 145 & 580 & 7×7×1 \\
A3-type AFM &  &  &  &  \\
FM   & M$_{Mn}$ = +1, M$_{Fe}$ = +1 & 80 & 320 & 7×7×1 \\
BLs & M$_{Mn}$-$_\text{top}$ = ±1, M$_{Fe}$-$_\text{bottom}$ = ±1 & 130 & 520 & 5×5×1 \\
\bottomrule
\end{tabular}
\end{table*}

\section{Results and discussion}

\subsection{Structural statement}
First, we optimized each monolayer to obtain the relaxed lattice parameters and atomic positions that serve as the structural baseline for the BL models. MSi$_2$N$_4$ (M = Fe, Mn) was relaxed in its hexagonal primitive unit cell (pcell) until the minimum-energy lattice constants and atomic coordinates were achieved. For Fe and Mn, the equilibrium in-plane lattice constants are $a=b$ $2.8698$ \AA \;and $2.847$ \AA, respectively. These calculations were performed without SOC. Our values are consistent with prior reports \cite{Ren2023MX2Y4} and are summarized in Table~\ref{tab:table1}.

The 2H-$\alpha$ phase examined in this work belongs to the D$_{3h}$ point group (Schoenflies notation), or equivalently the nonsymmorphic P$\bar{6}$m2 space group (Hermann-Mauguin notation). This symmetry class contains a horizontal mirror plane ($\sigma_h$), three vertical mirror planes ($\sigma_v$), three twofold rotation axes (C$_2$), and a sixfold rotation axis (C$_6$), and is non-centrosymmetric.
This symmetry class underlies several of the magnetic features discussed later. In particular, once SOC is included, MSN and FSN can break time-reversal symmetry and reduce the effective symmetry of the monolayers (MLs), a point we return to in the section on the primitive magnetic ground state. 

\begin{table*}[t]
\centering
\caption{Relaxed lattice constants, interlayer distance, magnetization, and MAE.}
\label{tab:table1}
\begin{tabular}{lccccc}
\toprule
\textbf{2D system} & \textbf{$a=b$ (\AA)} & \textbf{$d$ (\AA)} & \textbf{$M_{\rm tot}$ ($\mu_B$)} &
\textbf{$M_{\rm TM}/M_{\rm (TM-N)}$ ($\mu_B$)} & \textbf{MAE ($\mu$Ry)} \\
\midrule
FeSi$_2$N$_4$ & 2.8698 & -- & 3.61 & -- & 49.2 \\
MnSi$_2$N$_4$ & 2.8470 & -- & 3.199 & -- & $-57.21$ \\
\midrule
\multirow{2}{*}{BL}
& \multirow{2}{*}{2.851}
& \multirow{2}{*}{3.108}
& \multirow{2}{*}{8.17}
& 3.02 / $-0.192$ & \multirow{2}{*}{$-1\times10^{-7}$} \\
& & & & 3.50 / 0.026 & \\
\bottomrule
\end{tabular}
\end{table*}

Table~\ref{tab:table2} summarizes the optimized bond lengths. The transition-metal-nitrogen (TM-N) distances in the Fe, and Mn monolayers are $3.592/2.071$, $3.495/1.995$, and $3.483/1.989$~\AA, respectively. 

\begin{figure*}[htbp]
    \centering
    \includegraphics[width=0.8\textwidth]{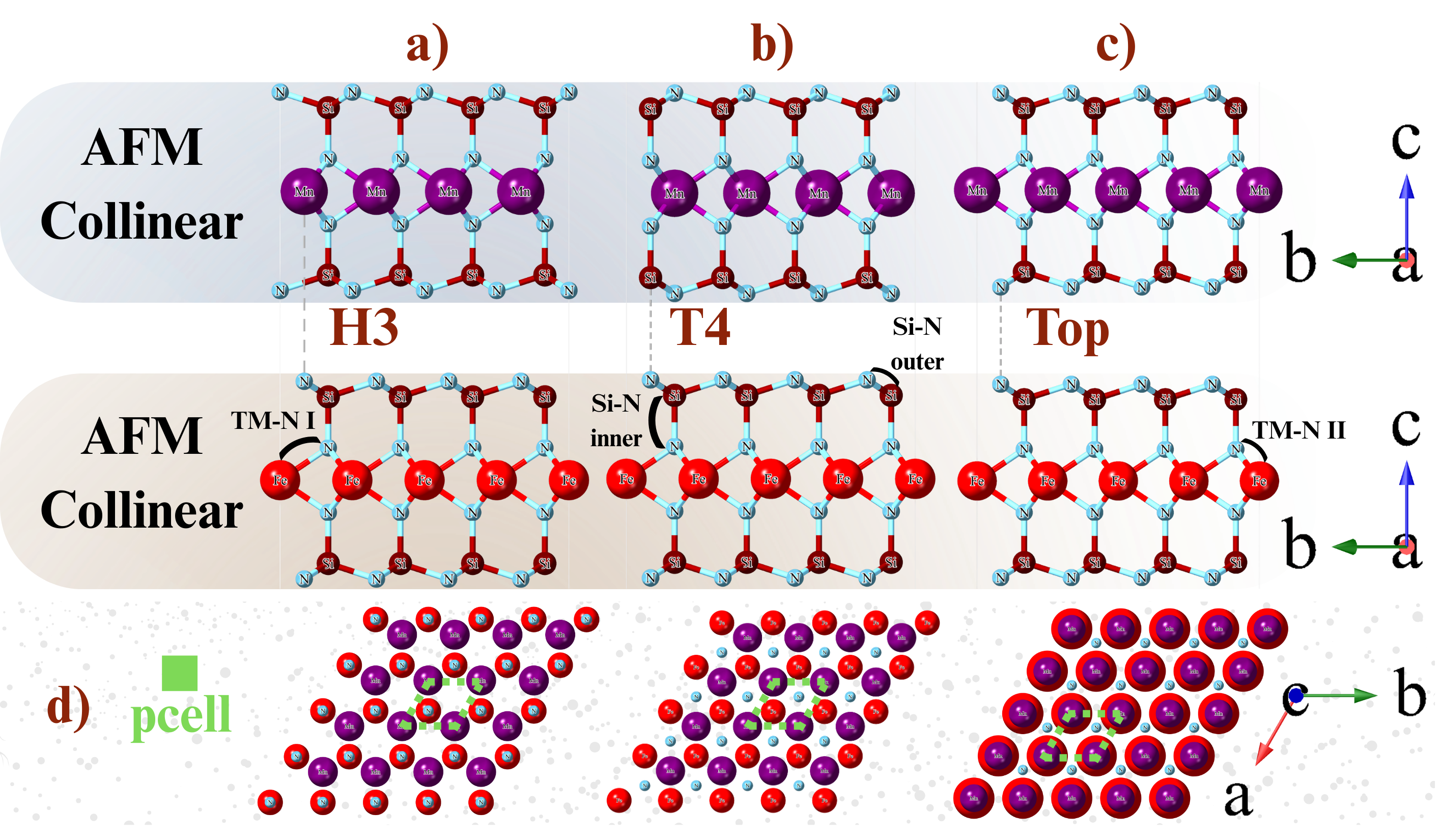}
    \caption{Side-view of the 2x2 \texttt{BLs} system showing the stacking a) H3, b) T4, c) Top where grey dashed lines point out high-symmetric interactions TM-N’s. Structural sections d) display their top-view above, excluding silicons and non-interfacial nitrogens, separately. Pcells are labeled in green.}
    \label{fig:modelvdWHS6}
\end{figure*}

To establish the structural configurations used later in our magnetic modeling, we considered all BL stackings compatible with the C$_6$ symmetry of these non-centrosymmetric MLs. In this work, the BLs correspond to MSN/FSN. Following the structural motifs explored in prior studies \cite{PedrozaRojas2025, BernalWalker1997}, we examined three representative registry types. In the H3 (Hexagonal-3) arrangement [Fig.~\ref{fig:modelvdWHS6}(a)], the transition metal (TM) atoms of one layer sit directly above the uppermost atomic plane of the adjacent layer. The T4 configuration [Fig.~\ref{fig:modelvdWHS6}(b)] corresponds to a registry in which the top nitrogen atoms align across layers, a geometry that can also arise at grain boundaries due to local structural distortions \cite{VanDerZande2013}. A third configuration, labeled Top [Fig.~\ref{fig:modelvdWHS6}(c)], places the surface nitrogen atoms in vertical alignment. Figure~\ref{fig:modelvdWHS6}(d) provides a top view of these registries, emphasizing the interlayer atomic alignments.

All structures were fully relaxed in $1\times1$ BLs (with TM positions fixed) to allow for possible local reconstruction. We then scanned the interlayer separation ranging from $\sim 10$~Å down to the smallest non-bonding distance to map the potential energy surface (PES). In every case, its minimum corresponded to an H3-like stacking, as shown in Fig.~\ref{fig:vdWHS_potential}, indicating that this registry is consistently the most stable non-covalently bonded configuration. 

Additionally, strain can induce artificial proximity effects~\cite{Zutic2019Proximitized}, potentially altering lattice constants, bond lengths, and even the resulting magnetic order. As shown in Table~\ref{tab:table1}, however, the lattice constants of the BL differ by only $\sim 0.246\%$ from those of the isolated MLs. Such a small mismatch places the system well within the regime where strain-induced proximity effects can be safely neglected.

\begin{table}[t]
\centering
\caption{Transition metal–inner nitrogen (TM–iN) I/II, silicon–inner nitrogen (Si–iN), and silicon–outer nitrogen (Si–oN) bonds.}
\label{tab:table2}
\begin{tabular}{lcccc}
\toprule
\textbf{2D system} &
\textbf{TM-iN I (\AA)} &
\textbf{TM-iN II (\AA)} &
\textbf{Si-iN (\AA)} &
\textbf{Si-oN (\AA)} \\
\midrule
FeSi$_2$N$_4$ & 3.495 & 1.995 & 1.752 & 1.739 \\
MnSi$_2$N$_4$ & 3.483 & 1.989 & 1.751 & 1.735 \\
\midrule
\multicolumn{5}{l}{\textbf{BL}} \\
MnSi$_2$N$_4$ & 3.335 & 1.990 & 1.745 & 1.730 \\
FeSi$_2$N$_4$ & 3.473 & 1.984 & 1.746 & 1.730 \\
\bottomrule
\end{tabular}
\end{table}

\begin{figure*}[htbp]
    \centering
    \includegraphics[width=0.7\textwidth]{Lattice_constantVSInterlayerDistance_HTMF_2025.png}
    \caption{PES in the 1x1 BLs, where $d$ is the interlayer distance.}
    \label{fig:vdWHS_potential}
\end{figure*}

\subsection{Primitive magnetic ground state}
The magnetic properties of the constituent MLs play a central role in determining the ground-state behavior of the resulting vdWH. Because these layers exhibit distinct electronic and magnetic characteristics, it is necessary to examine how the interfacial environment in Bernal-like stackings modifies their interactions.

We first note that the net magnetizations of the MLs do not change abruptly under stacking, which supports our decision to neglect SOC in the Heisenberg modeling. This does not imply the absence of intrinsic SOC; rather, its contribution remains secondary to the dominant exchange processes. When SOC is included at the DFT level, the systems exhibit metallic behavior consistent with prior reports~\cite{Zhao2023}. In this regime, Fe-containing MLs carry a larger magnetic moment ($3.61\mu_B$), indicating an increased density of spin-polarized conduction carriers (see Table~\ref{tab:table1}).

Regarding the inquiry of how interlayer interactions modify the spin structure, we investigated it by examining the magnetic behavior of the BLs. As summarized in Table~\ref{tab:table4}, the total energies of the various interlayer magnetic alignments all lie within $\sim 3$~meV of one another, indicating that the BL remains structurally stable regardless of the relative spin configuration~\cite{Tung2016}. Among these, the most energetically favorable ordering is $M_{B_{\uparrow}}^{A_{\uparrow}}$, followed sequentially by $M_{B_{\downarrow}}^{A_{\uparrow}}$ and $M_{B_{\uparrow}}^{A_{\downarrow}}$. The small energy differences between these states reveal a delicate balance of attractive and repulsive interlayer exchange channels, which in turn enable rapid vertical (interlayer) magnetic phase changes. The resulting interlayer coupling slightly reduces the local magnetic moments--by roughly $0.1~\mu_B$--relative to the isolated monolayers.
\begin{table}[t]
\centering
\caption{Relative energies for different magnetic alignments of the bilayer (BL). Energies are given relative to the lowest-energy configuration.}
\label{tab:table4}

\begin{tabular}{lccc}
\toprule
Configuration
& $M_{(B\downarrow)}^{(A\uparrow)}$
& $M_{(B\uparrow)}^{(A\downarrow)}$
& $M_{(B\uparrow)}^{(A\uparrow)}$ \\
\midrule
Energy (mRy) & 0.13 & 0.21 & 0 \\
\bottomrule
\end{tabular}

\vspace{0.15em} 

\begin{tabular}{lcc}
\toprule
Layer & Mn ($\mu_B$) & Fe ($\mu_B$) \\
\midrule
BL & 3.02 & 3.50 \\
\bottomrule
\end{tabular}
\end{table}

Switchable magnetocrystalline anisotropy emerges in these magnetic compounds under charge doping, which perturbs the in-plane $D_{3h}$ orbital environment. Although this mechanism does not induce the type of strongly mixed relativistic spinor structure generated by SOC in the (001) direction, it does break the in-plane symmetry and lifts the degeneracy between in-plane and out-of-plane spin orientations. Such symmetry lowering can also allow spin–quantum–Hall–like responses in principle.

To quantify the magnetic easy axis, we evaluate the Single-Ion Anisotropies (SIAs) extracted from the MAE calculations. FSN prefers an out-of-plane (OOP) easy axis, whereas MSN favors an in-plane (IP) alignment--consistent with the nearly isotropic crystal field experienced by Mn in this environment. The corresponding MAE values are $-669.401~\mu\mathrm{eV}$ ($-0.669~\mathrm{meV}$) for OOP alignment in FSN and $+778.382~\mu\mathrm{eV}$ ($+0.778~\mathrm{meV}$) for IP alignment in MSN. These give SIA energies of $238.82~\mu\mathrm{eV}$/Fe and $-261.7~\mu\mathrm{eV}$/Mn. The magnitude of the Mn SIA, particularly when compared with the reported $-365~\mu\mathrm{eV}$/Mn value, indicates that relativistic corrections and the exchange–correlation damping term play a significant role. It is well known that the Generalized Gradient Approximation tends to overestimate magnetic moments, while the Local Density Approximation underestimates exchange constants. Even so, our anisotropy values remain consistent with prior work and compare unfavorably against related Fe/Mn-based two-dimensional compounds. For example, Fe$_3$GeTe$_2$ monolayers exhibit an OOP MAE of $2.68~\mathrm{meV}$~\cite{PhysRevB.105.014437}, substantially larger than our FSN result, it needs at least two single monolayers to conserve Weyl topological nodes\cite{roemer_unraveling_2024} though. While the IP anisotropy and SIA reported for MnIn$_2$Se$_3$I$_2$ ($\sim12.667~\mathrm{meV}$ and $931~\mu\mathrm{eV}$)~\cite{hnm7-xz31} are larger, but it arises in a far more structurally complex and defect-prone material. 
\begin{figure}[h]
    \centering
    \includegraphics[width=0.5\textwidth]{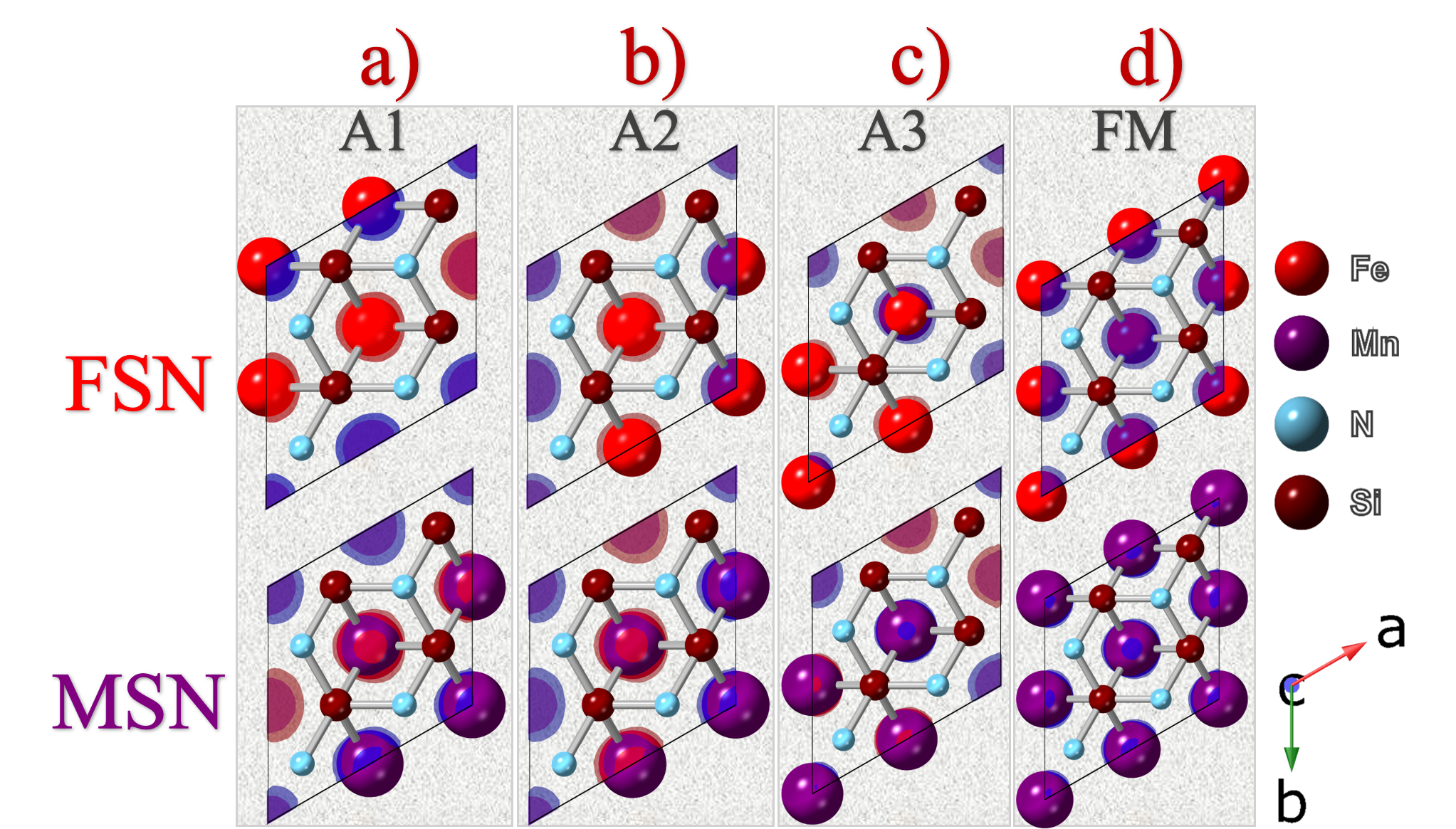}
    \caption{Spin isosurfaces for MSN and FSN in a 2$\times$2 periodic arrangement. Blue and red lobes represent the accumulation of spin-up and spin-down probability densities, respectively, for each spin channel.}
    \label{fig:MLs_magneticorders}
\end{figure}
When on-site Coulomb interactions are incorporated through $U_{\mathrm{eff}}$, the MLs display AFM tendencies in accordance with Hund’s rules, consistent with earlier studies~\cite{Chen2022, An2023}. Because the TM atoms occupy edge-sharing environments in our geometry, we reproduced the $2\times2$ magnetic configurations shown in Fig.~\ref{fig:MLs_magneticorders} to verify the expected ordering. Although the magnetic moments differ slightly $-2.843/-2.862\mu_B$ for Mn and $\pm3.318\mu_B$ for Fe--this variation is consistent with a more accurate treatment of electronic correlations, including Becke–Johnson damping and tighter convergence thresholds \cite{grimme2011}. Thus, our calculations remain in agreement with the predicted ML magnetic ground states while providing improved numerical precision.

\begin{table*}[t]
\centering
\caption{Relative energy, magnetic moments, and bond lengths for MSN and FSN under various magnetic configurations. Energies are relative to the ground state (A1).}
\label{tab:merged_magnetic_data}
\begin{tabular}{llcccccc}
\toprule
\textbf{System} &
\textbf{Order} &
\textbf{Rel. Energy} &
\textbf{$\mu_B$ per TM} &
\multicolumn{3}{c}{\textbf{Bonds (\AA)}} \\
\cmidrule(lr){5-7}
& & \textbf{(mRy)} & &
\textbf{TM--N I} &
\textbf{TM--N II} &
\textbf{Si--iN/oN} \\
\midrule
\multirow{4}{*}{MnSi$_2$N$_4$}
 & A1 & 0.000  & $+2.843,\,-2.862$ & 3.476 & 1.984 & 1.747/1.732 \\
 & A2 & 9.567  & $+2.843,\,-2.862$ & 3.476 & 1.984 & 1.745/1.729 \\
 & A3 & 9.571  & $+2.843,\,-2.862$ & 3.467 & 1.983 & 1.745/1.729 \\
 & FM & 83.277 & $\pm 2.831$        & 3.471 & 1.987 & 1.745/1.729 \\
\midrule
\multirow{4}{*}{FeSi$_2$N$_4$}
 & A1 & 0.000  & $\pm 3.436$ & 3.482 & 1.980 & 1.744/1.735 \\
 & A2 & 1.107  & $\pm 3.436$ & 3.480 & 1.979 & 1.745/1.734 \\
 & A3 & 1.107  & $\pm 3.436$ & 3.473 & 1.980 & 1.745/1.735 \\
 & FM & 27.882 & $\pm 3.318$ & 3.478 & 1.984 & 1.746/1.733 \\
\bottomrule
\end{tabular}
\end{table*}

On the other hand, when the Hubbard $U$ and Hund’s coupling $J$ are not included through Dudarev’s correction for the Fe–$d$ and Mn–$d$ orbitals in FSN and MSN \cite{Wu2022MoSi2N4}, the MLs remain metallic in the spin-polarized channel. This contrasts with the primitive (i.e., $U\neq0$) description, where the same systems appear half-metallic due to the effective Coulomb repulsion within the spinor manifold, which stabilizes sextet and octet spin multiplicities. A closer inspection of the separate roles of $U$ and $J$ further reveals that the coupling parameter $J$ influences the Dirac-like valence band in MSN, modifying the relative splittings near the Fermi level.

In the AFM phases of MSN and FSN, the fully relaxed $2\times2$ supercells show that the Hubbard interaction changes the equilibrium volume (introducing local structural disorder)--consistent with previous studies \cite{Zhao2023,Tian2022ElectronicPropertiesMnSi2N4}. This indicates that the energetics of the AFM configurations depend sensitively on crystal-field splitting, whereas the FM phase remains comparatively unaffected, reflecting the presence of Kramers degeneracy and broken time-reversal symmetry only in the latter. As a result, the MLs adopt the magnetic space group P$_B$mn2$_1$ (type-IV), whose corresponding spin group is P$^1$m$^1$m$^1$2$^{-1}$. By symmetry, such a magnetic configuration cannot host an anomalous Hall effect. 

\subsubsection{Finite size in BLs}
All BL pcells were first relaxed while constraining the TM sites and allowing small displacements along the $z$-axis and in the $(100)/(010)$ planes. This initial relaxation produced a modest surface reconstruction, characterized mainly by slight OOP shifts and small changes in local bond lengths. The reconstructed geometries were then used as input for building $2\times2$ supercells, which were again relaxed under the same TM constraints but with fixed IP lattice parameters and restricted OOP motion. These supercells provided the IP periodic boundary conditions employed in the magnetic calculations, which were performed self-consistently using the rotationally invariant Dudarev DFT$+U$ approach incorporating both Hubbard $U$ and Hund $J$.

The Si–N sublayers in MA$_2$Z$_4$ play an essential structural and electronic role in stabilizing the magnetic BLs. Nitrogen contributes lone-pair $p$ electrons that participate in mixed ionic–covalent bonding, forming a mechanically rigid and electronically active Si-N framework on either side of the TM plane. This bonding character allows charge redistribution across the BL without introducing strong covalency through the van der Waals gap, thereby maintaining well-defined local TM moments, as summarized in Table~\ref{tab:table6}.

To fully characterize possible BL magnetic configurations, all symmetry-allowed combinations arising from the non-centrosymmetric MLs were examined. In total, eight distinct interlayer magnetic couplings were constructed and evaluated, consistent with the geometries illustrated in Fig.~\ref{fig:AFMordersvdWHS6} of the Atomistic Exchange Model section.

Surface reconstruction effects, driven by the local changes in bond lengths, can be quantified directly from the DFT$+U$ results. Relative to the pcells, the AFM-coupled $2\times2$ BLs exhibit TM moment variations of approximately $+0.327$/$+0.308~\mu_B$ for Mn and $\pm0.064~\mu_B$ for Fe. For FM-coupled BLs, the corresponding changes increase to $\pm0.539~\mu_B$ and $\pm0.285~\mu_B$, respectively. These magnetic reorganizations correlate with small but systematic modifications of the Si-iN and Si-oN bond lengths in MSN (about $-0.173\%$ for FM and $-0.400\%$ for AFM), and similar changes for Mn-N(I/II) bonds ($-0.197\%$ for FM and $-0.146\%$ for AFM).

For FSN, the Si-iN/oN bond variations are $-0.337\%$ (FM) and $-0.400\%$ (AFM), while Fe-N(I/II) bonds shift by $-0.197\%$ (FM) and $-0.823\%$ (AFM). These distortions remain far from those required for a Jahn-Teller instability: the TM-N(I/II) octahedra do not distort toward a tetrahedral environment. However, the observed pattern of small bond-dependent moment changes suggests the possible emergence of nanoscale magnetic domain walls, similar to those recently identified in AFM BL materials such as CrSBr~\cite{hiendrich,Zur2023,tschudin_imaging_2024,polymorphism_CSB}.

\begin{table*}[ht]
\centering
\caption{Relationship of the various magnetic orders represented in Figure~\ref{fig:AFMordersvdWHS6}.}
\label{tab:table6}
\begin{tabular}{llllllll}
\hline
\textbf{Order} & \textbf{Rel. Energy} & \textbf{M$_{Mn}$} & \textbf{M$_{Fe}$} & \textbf{M$_{Tot}$} & \multicolumn{3}{c}{\textbf{Distances (\AA)}} \\
\cline{6-8}
& \textbf{(mRy)} & \textbf{($\mu_B$)} & \textbf{($\mu_B$)} & \textbf{($\mu_B$)} & Mn-iN I/II & Fe-iN I/II  & d \\
\hline
FM & 63.64 & 3.17 & 3.60 & 28 & 2.030/2.030 & 2.050/3.512 & 2.915 \\
$A1$ & 64.09 & -3.17 & 3.60 & -4.12 & 2.030/2.030 & 2.050/3.512 & 2.915 \\
$A2$ & 64.13 & 3.17 & -3.60 & -4.12 & 2.030/2.030 & 2.050/3.512 & 2.92 \\
$C1$ & 0.051 & $\pm$2.97 & $\pm$3.54 & 0 & 2.033/2.019 & 2.052/3.495 & 2.96 \\
$G1$ & 0 & $\pm$2.97 & $\pm$3.54 & 0 & 2.033/2.019 & 2.052/3.495 & 2.96 \\
$C2$  & 0 & $\pm$2.97 & $\pm$3.54 & 0 & 2.035/2.032 & 2.049/3.511 & 2.92 \\
$G2$ & 0.782 & $\pm$2.97 & $\pm$3.54 & 0 & 2.035/2.032 & 2.049/3.511 & 2.92 \\
$C3$ & 0.782 & $\pm$2.97 & $\pm$3.54 & 0 & 2.021/2.023 & 2.048/3.515 & 2.945 \\
$G3$ & 0.884 & $\pm$2.97 & $\pm$3.54 & 0 & 2.021/2.023 & 2.048/3.515 & 2.945 \\
\\
\hline
\end{tabular}
\end{table*}

\subsection{Atomistic exchange model}
To extract the effective exchange coupling constants, we adopt an isotropic Heisenberg description of the localized spin moments and neglect on-site magnetocrystalline anisotropy energy (MAE) \cite{choi2019}. The general form of the spin Hamiltonian is
\begin{equation}\label{FullHam}
    \mathcal{H} = \sum_{i<j} \hat{\vec{S}}_i \cdot \mathrm{J}_{ij} \cdot \hat{\vec{S}}_j,
\end{equation}
where $\mathrm{J}_{ij}$ denotes the full exchange tensor coupling sites $i$ and $j$ \cite{facemyer2023, facemyer2025}.

The MAE is omitted for two reasons.
First, the energy differences used to extract $J_a$, $J_d$, and $J_{2a}$ come from comparing collinear FM and AFM configurations; the MAE contributes only an overall shift and cancels exactly between these states. Second, the bilayer MAE obtained from DFT is extremely small (of order $10^{-7}$~$\mu$Ry), meaning it does not influence the low-energy spectrum or the correlation observables obtained from exact diagonalization. In other words, the system is far from a spin-reorientation regime, and the isotropic Heisenberg form accurately captures the relevant physics \cite{Xiang2013}.

\begin{figure*}[htbp]
    \centering
    \includegraphics[width=1.0\textwidth]{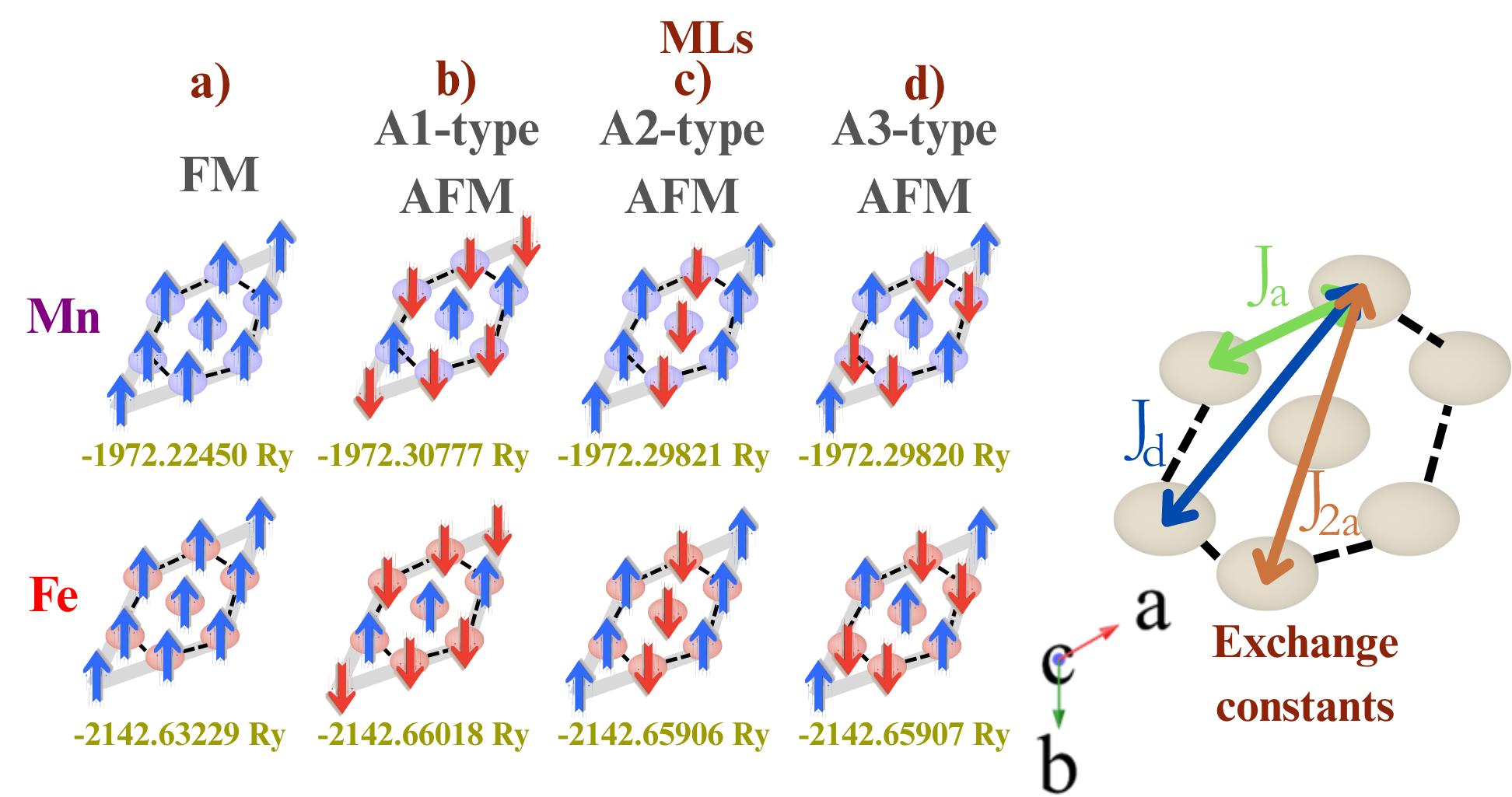}
    \caption{Representative model of nearest ($J_a$, exchange), next ($J_d$, double-exchange), and third ($J_{2a}$, super-exchange)-second-nearest neighbors intralayer terms for AFM/FM alignments: MSN and FSN in 2x2 periodicity.}
    \label{fig:AFMorders}
\end{figure*}

We begin with the isolated monolayers (MLs), where the magnetic ions form nearly triangular in-plane arrays. Each ML is described using the three dominant intralayer exchange paths illustrated in Fig.~\ref{fig:AFMorders}: nearest neighbors ($J_a$), diagonal next-nearest neighbors ($J_d$), and second-nearest neighbors separated by twice the lattice constant ($J_{2a}$). This yields the single-layer effective model
\begin{equation}\label{mlHam}
    \mathcal{H}_0 = J_a \sum_{\langle ij \rangle} \hat{\vec{S}}_i \cdot \hat{\vec{S}}_j + 
              J_d \sum_{\langle\langle ij \rangle\rangle} \hat{\vec{S}}_i \cdot \hat{\vec{S}}_j + 
              J_{2a} \sum_{\langle\langle\langle ij \rangle\rangle\rangle} \hat{\vec{S}}_i \cdot \hat{\vec{S}}_j,
\end{equation}
where $J_{ij}<0$ favors FM alignment and $J_{ij}>0$ favors AFM alignment. The angled-bracket notation in the summation limits specifies the degree of nearness between lattice sites.

Using relaxation energies obtained from DFT for several magnetic configurations within a $2\times 2$ periodic supercell (also shown in Fig.~\ref{fig:AFMorders}), we extract
\begin{align}
    E_{F}  &= 2S_z^2 \big( 13J_a + 11J_d + 12J_{2a} \big), \nonumber \\
    E_{A1} &= 2S_z^2 \big( -5J_d + J_{2a} \big), \nonumber \\
    E_{A2} &= 2S_z^2 \big( -3J_a + 3J_d - 4J_{2a} \big), \nonumber \\
    E_{A3} &= 2S_z^2 \big( J_a - 5J_d + 3J_{2a} \big),
\end{align}
with the corresponding energy differences
\begin{align}
    \Delta E_1 &= E_{F} - E_{A1} = 2S_z^2 \big( 13J_a + 16J_d + 11J_{2a} \big), \nonumber \\
    \Delta E_2 &= E_{F} - E_{A2} = 16S_z^2 \big( 2J_a + J_d + 2J_{2a} \big), \nonumber \\
    \Delta E_3 &= E_{F} - E_{A3} = 2S_z^2 \big( 12J_a + 16J_d + 9J_{2a} \big). \label{DeltaEintra}
\end{align}

For Mn ($S=5/2$), this yields $J_a \approx -4.65$ meV ($\approx -53.94$ K), $J_d \approx +4.27$ meV ($\approx +49.54$ K), and $J_{2a} \approx +7.53$ meV ($\approx +87.35$ K). For Fe ($S=7/2$), $J_a \approx +0.915$ meV ($\approx +10.61$ K), $J_d \approx +0.327$ meV ($\approx +3.79$ K), and $J_{2a} \approx -0.149$ meV ($\approx -1.73$ K). Even though Mn in MSN is a single monolayer of MA$_2$Z$_4$, its intraexchange constant is much greater than that of bulk MnBi$_2$Te$_4$ of +0.4 meV \cite{Chen2024} and +0.77 meV in undistorted MnPSe$_3$\cite{PhysRevB.103.024414}. This is one example of why the result is a breakthrough on its own. When comparing FSN exchanges with greater values in FM and AFM, favoring exchange values of 3.27 and -0.44 meV, respectively, in the bulk of Fe$_3$GeTe$_2$\cite{liu_layer-dependent_2022}, whose thickness is 8.6 \AA, and its uncompensated Fe$^{2+}$/Fe$^{3+}$ itinerant ferromagnetic electrons, the result is significant. Similarly, FeCl$_2$ has a reported exchange constant of $J=+0.7$ meV at its Curie temperature T$_C=13$ K \cite{Yang2020FeCl2Strain}. At $T=0$ K, this value is expected to be smaller, since DFT calculations include the dipole-dipole (zeroth-order) contribution that reduces the effective exchange.
 
\subsubsection{Bilayer coupling}
\begin{figure*}[htbp]
    \centering
    \includegraphics[width=1.0\textwidth]{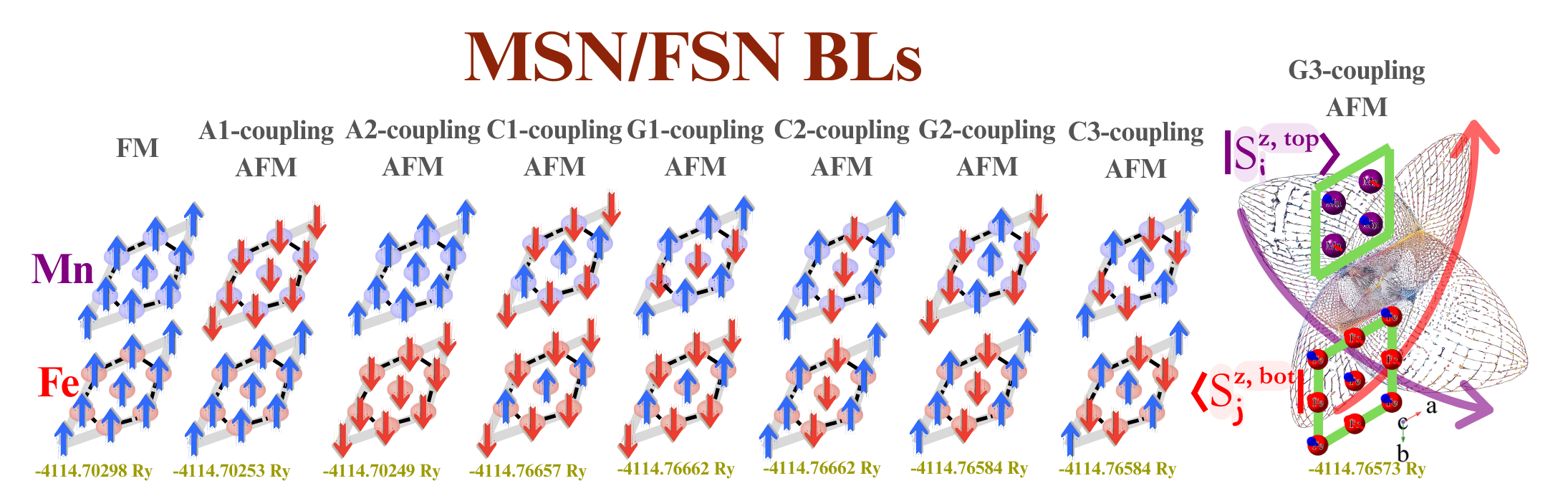}
    \caption{Representative model of FM, A-, C- and, G-couplings: MnSi$_2$N$_4$/FeSi$_2$N$_4$ – stacking H3 operated in the Heisenberg hamiltonian. It is worth noting that the BL allows four magnetic alignments more. A at G3-coupling}
    \label{fig:AFMordersvdWHS6}
\end{figure*}

We next couple the two MLs (Fig.~\ref{fig:AFMordersvdWHS6}) through interlayer Heisenberg exchange. Each site $i$ in the top layer couples both to the vertically aligned site $j=i$ in the bottom layer (nearest interlayer neighbor) and to the surrounding sites sharing the same in-plane coordinate up to diagonal distance (next-nearest interlayer neighbors). The resulting bilayer Hamiltonian is
\begin{equation}\label{bl_ham}
\mathcal{H} = \mathcal{H}_0 + \sum^{N}_{ij} J_{ij} \left( S_i^{z,\mathrm{top}} S_j^{z,\mathrm{bot}} + \tfrac{1}{2} (S_i^{+,\mathrm{top}} S_j^{-,\mathrm{bot}} + S_i^{-,\mathrm{top}} S_j^{+,\mathrm{bot}}) \right),
\end{equation}
with $N$ sites in each layer and the interlayer neighbor shells defined as $J_{\langle ij \rangle}$ (same site) and $J_{\langle\langle ij \rangle\rangle}$ (adjacent and diagonal sites).

Energy differences relative to the FM reference configuration yield
\begin{align}
\Delta E_4 &= E_{F} - E_{A1} = 157.5 J_{\langle ij \rangle} + 218.75 J_{\langle \langle ij \rangle \rangle}, \nonumber \\
\Delta E_5 &= E_{F} - E_{A2} = 350 J_{\langle \langle ij \rangle \rangle}. \label{DeltaEinter}
\end{align}
These give $J_{\langle ij \rangle} \approx +5.46$ meV ($\approx 63.34$ K) and $J_{\langle \langle ij \rangle \rangle} \approx -0.018$ meV ($\approx -0.21$ K).

\begin{figure*}[t]
    \centering
    \begin{subfigure}{0.48\textwidth}
        \centering
        \includegraphics[width=\linewidth]{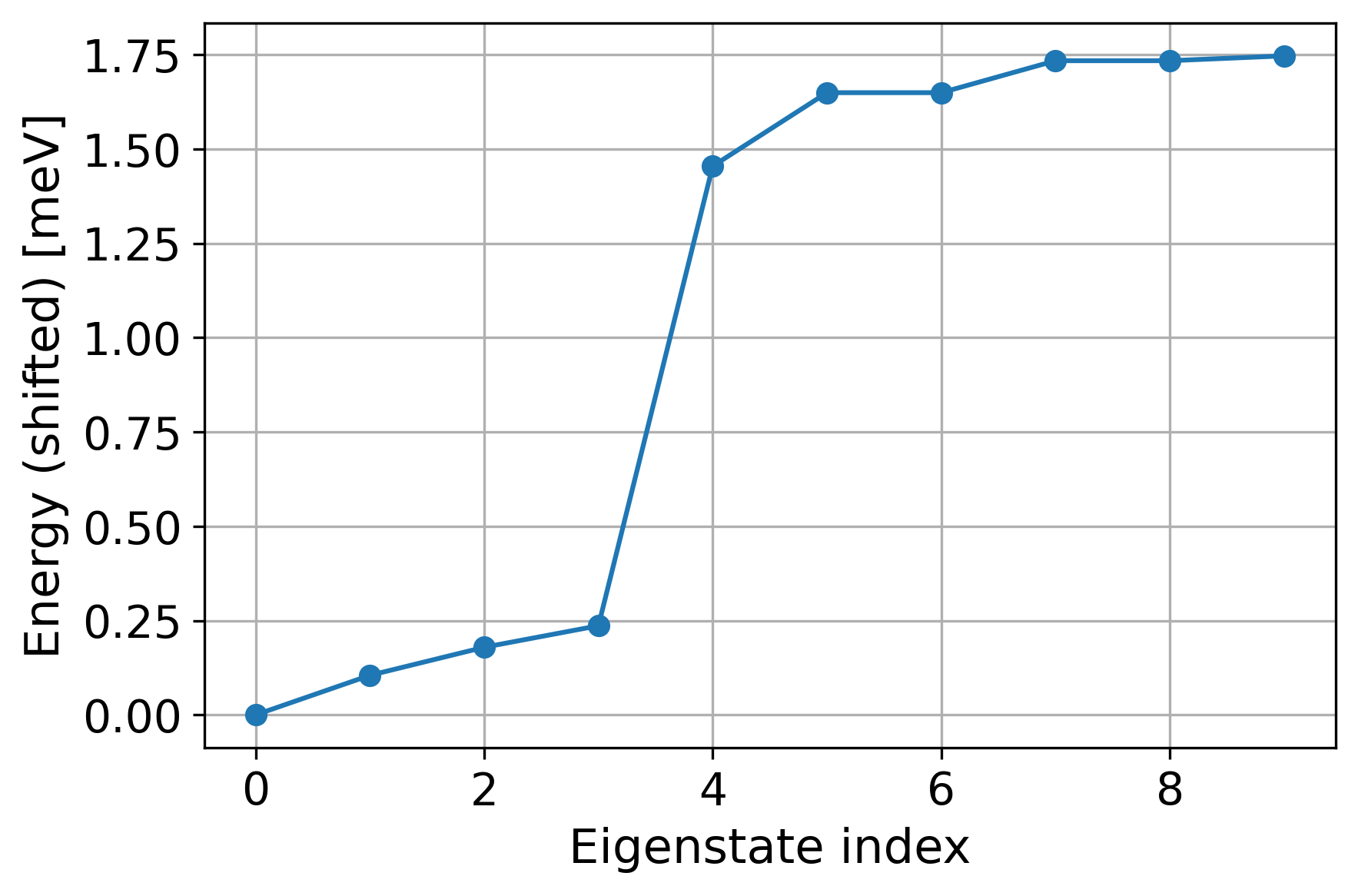}
        \caption{}
        \label{fig:SpectrumBL}
    \end{subfigure}
    \hfill
    \begin{subfigure}{0.48\textwidth}
        \centering
        \includegraphics[width=\linewidth]{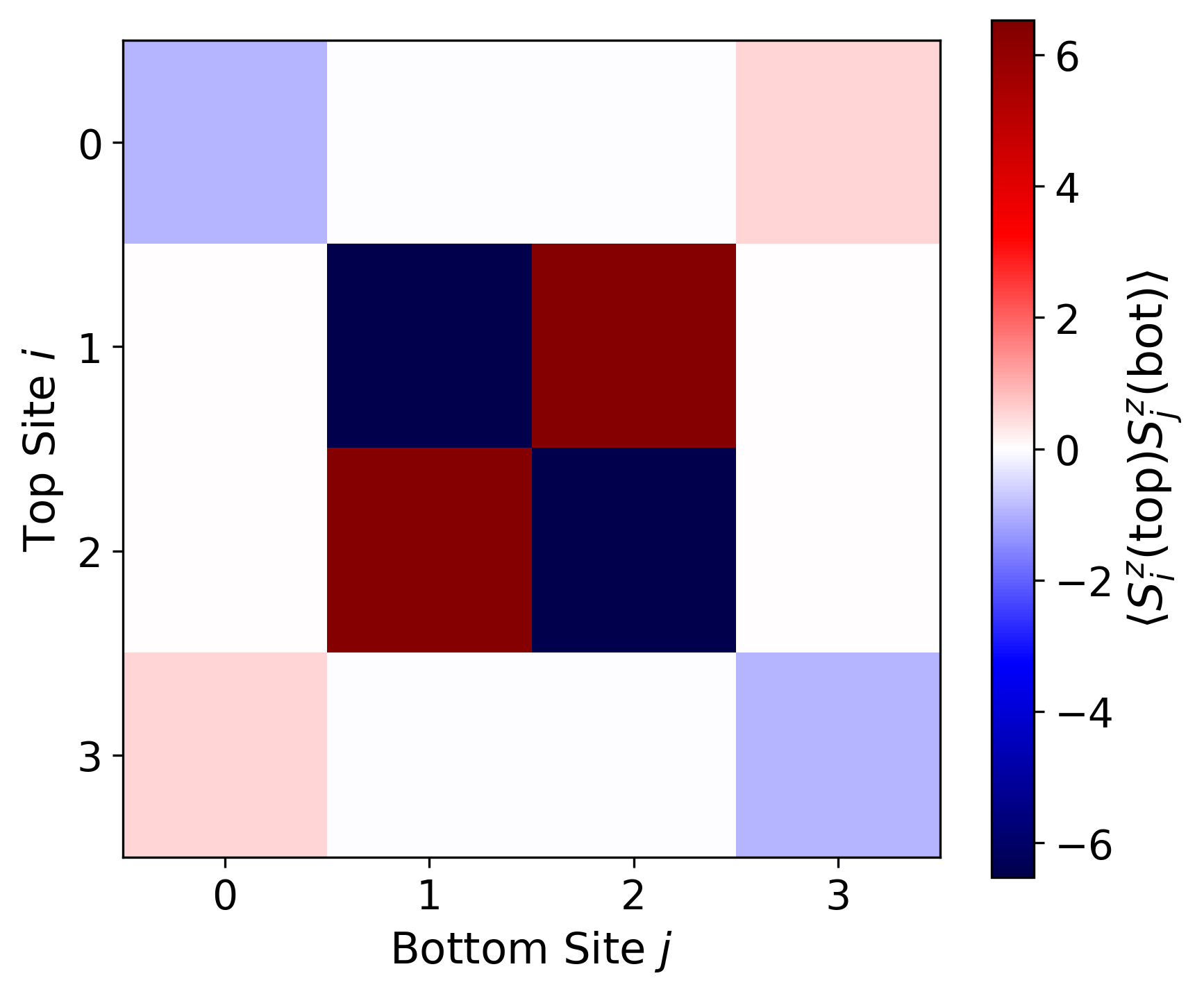}
        \caption{}
        \label{fig:CorrBL}
    \end{subfigure}

    \caption{
    (a) Low-lying many-body spectrum of the MnSi$_2$N$_4$/FeSi$_2$N$_4$ bilayer Heisenberg model with toroidal boundary conditions, using intralayer and interlayer exchange parameters extracted from DFT.
    (b) Cross-layer spin--spin correlation matrix $\langle S^{z,\mathrm{top}}_i S^{z,\mathrm{bot}}_j\rangle_{\mathrm{GS}}$. Red (blue) denotes ferromagnetic (antiferromagnetic) correlations between site $i$ in the top layer and site $j$ in the bottom layer. The alternating pattern among interior sites is consistent with the G2- and G3-type interlayer antiferromagnetic configurations shown in Fig.~\ref{fig:AFMordersvdWHS6}.
    }
    \label{fig:SpectrumAndCorr}
\end{figure*}

Exact diagonalization of the bilayer Heisenberg Hamiltonian, Eq.~\ref{bl_ham}, using the intralayer and interlayer exchanges extracted above, yields the low-lying spectrum shown in Fig.~\ref{fig:SpectrumAndCorr}(a). The nondegenerate ground state in the $m=0$ sector is separated from the first excited state by a gap of order $0.13$~meV, while a group of higher excitations clusters between $\sim 1.5$ and $1.75$~meV. To probe the real-space spin structure encoded in this ground state, we evaluate the cross-layer correlations $\langle S^{z,\mathrm{top}}_i S^{z,\mathrm{bot}}_j\rangle_{\mathrm{GS}}$, shown in Fig.~\ref{fig:SpectrumAndCorr}(b). White entries correspond to site pairs whose spin correlations vanish. This indicates that their contributions cancel across the quantum ground state superposition, either due to symmetry, frustration, or competing FM/AFM exchange paths. These pairs do not participate in the dominant interlayer ordering pattern. The alternating ferro- and antiferromagnetic correlations between the interior lattice sites reproduce the same ordering pattern as the G2- and G3-type interlayer antiferromagnetic configurations illustrated in Fig.~\ref{fig:AFMordersvdWHS6}. This establishes that the G-type arrangements dominate the low-energy spin texture of the MnSi$_2$N$_4$/FeSi$_2$N$_4$ bilayer and are the natural continuation of the monolayer-derived exchange interactions once the layers are coupled.

In summary, A1-type AFM for Mn and Fe MLs has the potential to modify the interaction in its entirety due to the finite size, given the existence of longer-range interactions. It has been demonstrated that both systems conserve $\hat{P}\hat{T}$ in the stripe AFM, thereby fostering the possibility of facile magnetic phase transitions (see reference \cite{doi:10.1126/sciadv.aaz8809, Zhou2025AFMtoAM} for details). Magnetic transition from intralayer exchange A1-type in the isolated materials to A2-type and A3-type in the G2 and G3 couplings corroborates that a research gap remains concerning the absence of the necessity of SOC and how the conservation of flat bands and tilted cones in the pcells of A1-type MLs changes. This transformation of their ground state renders them not only superior to others but also exotic and non-trivial. Given the limitations imposed by the Mermin–Wagner theorem, the suppression of long-range antiferromagnetic order in the BL 2D regime leads to the formation of locally uncompensated spin regions. In the presence of external magnetic fields and proximity-induced itinerant carriers, these regions can act as effective internal exchange fields, leading to Zeeman-split electronic Landau levels. This scenario naturally points to the coexistence of two local spin multiplicities in the bilayer stacks, reflecting spatially inhomogeneous antiferromagnetic compensation. 
 
\section{Conclusions}
In conclusion, the MA$_2$Z$_4$ bilayers studied here provide a compelling platform for exploring magnetic ordering and interlayer coupling in van der Waals systems. By combining first-principles calculations with an effective Heisenberg description and exact diagonalization, we demonstrate that the interlayer exchange interactions are non-perturbative and play a decisive role in shaping the magnetic ground state. Importantly, these interactions are comparable in magnitude to the dominant intralayer exchanges in at least one of the constituent layers, indicating that bilayer coupling cannot be treated as a weak correction to monolayer physics.

The robustness of the extracted exchange interactions across different supercell sizes and stacking configurations further indicates that the observed magnetic behavior is intrinsic to the bilayer geometry rather than a finite-size or symmetry artifact. While the present work focuses on static magnetic ordering and low-lying magnetic excitations, the resulting quantum correlation patterns highlight the importance of collective spin behavior beyond a purely single-particle picture.

These findings motivate future investigations into dynamical and transport phenomena in MA$_2$Z$_4$ heterostructures, where the interplay between strong exchange interactions, dimensionality, and symmetry may enable controllable spin configurations and antiferromagnetic functionality. In this context, a comprehensive understanding of spin dynamics should incorporate both localized magnetic excitations and collective spin-wave modes, particularly in regimes where antiferromagnetic correlations dominate.

\medskip
\textbf{Acknowledgements} \par 
The authors are grateful to Sergio E. Ulloa for insightful discussions on symmetry aspects of spin systems and the implications of spin--orbit coupling. The DFT calculations were carried out at the DGTIC-UNAM Supercomputing Center on the Xiuhcoatl node as part of the LANCAD-CINVESTAV-CGSTIC-65-2026 project. The first author is grateful to the Nanoscale and Quantum Phenomena Institute (NQPI), Ohio University (OU), and Prof. Ulloa for the financial support in a research visit.

\bibliographystyle{iopart-num} 
\bibliography{iop-references}

@article{grimme2016,
  title={A consistent and accurate ab initio parametrization of density functional dispersion correction (DFT-D) for the 94 elements H-Pu},
  author={Grimme, Stefan and Antony, Jens and Ehrlich, Stephan and Krieg, Helge},
  journal={The Journal of Chemical Physics},
  volume={132},
  number={15},
  pages={154104},
  year={2010},
  note={Often cited for D3. For BJ-damping specifically, cite: J. Comput. Chem. 32, 1456 (2011)}
}

@article{dudarev1998,
  title={Electron-energy-loss spectra and the structural stability of nickel oxide: An LSDA+U study},
  author={Dudarev, S. L. and Botton, G. A. and Savrasov, S. Y. and Humphreys, C. J. and Sutton, A. P.},
  journal={Phys. Rev. B},
  volume={57},
  pages={1505},
  year={1998},
  publisher={American Physical Society}
}

@article{Liechtenstein1995,
  title={Density-functional theory and strong interactions: Orbital ordering in Mott-Hubbard insulators},
  author={Liechtenstein, A. I. and Anisimov, V. I. and Zaanen, J.},
  journal={Phys. Rev. B},
  volume={52},
  pages={R5467},
  year={1995},
  publisher={American Physical Society}
}

@article{Wu2022MoSi2N4,
  title={Intrinsic ferromagnetism and high-temperature magnetic anisotropy in 2D MSi$_2$N$_4$ (M= Mn, Fe, Co, Ni) monolayers},
  author={Wu, Q. and others},
  journal={Applied Physics Letters},
  volume={120},
  pages={092406},
  year={2022}
}

@article{jia_spintronic_2025,
  title={Altermagnetic spintronics},
  author={Smejkal, Libor and Sinova, Jairo and Jungwirth, Tomas},
  journal={arXiv preprint arXiv:2508.09748},
  year={2025},
  note={Submitted Aug 2025. Alternatively, check ACS Nano 2025 for 'Spintronic Devices upon 2D Magnetic Materials'}
}

@article{liu2021,
author = {Liu, Lixin and Zhai, Tianyou},
title = {Wafer-scale vertical van der Waals heterostructures},
journal = {InfoMat},
volume = {3},
number = {1},
pages = {3-21},
doi = {https://doi.org/10.1002/inf2.12164},
year = {2021}
}

@article{39xc-2f3m,
  title = {Quantum-well resonance enhanced tunneling magnetoresistance effect in magnetic tunnel junctions with oxidized interface},
  author = {Jiang, L. N. and Chi, B. Y. and Han, X. F.},
  journal = {Phys. Rev. B},
  volume = {111},
  issue = {22},
  pages = {224422},
  numpages = {6},
  year = {2025},
  month = {Jun},
  publisher = {American Physical Society},
  doi = {10.1103/39xc-2f3m},
  url = {https://link.aps.org/doi/10.1103/39xc-2f3m}
}

@misc{zhong2024integrating,
  title        = {Integrating 2D Magnets for Quantum Devices: from Materials and Characterization to Future Technology},
  author       = {Zhong, Han and Plummer, Douglas Z. and Lu, Pengcheng and Li, Yang and Leger, Polina A. and Wu, Yingying},
  year         = {2024},
  eprint       = {2406.12136},
  archivePrefix = {arXiv},
  primaryClass = {cond-mat.mes-hall},
  url          = {https://arxiv.org/abs/2406.12136}
}

@article{PhysRevB.110.L220409,
  title = {Van der Waals spin-orbit torque antiferromagnetic memory},
  author = {Zhang, Lishu and Yuan, Zhengping and Yang, Jie and Zhou, Jun and Jiang, Yanyan and Li, Hui and Cai, Yongqing and Tsymbal, Evgeny Y. and Feng, Yuan Ping and Zhu, Zhifeng and Shen, Lei},
  journal = {Phys. Rev. B},
  volume = {110},
  issue = {22},
  pages = {L220409},
  numpages = {7},
  year = {2024},
  month = {Dec},
  publisher = {American Physical Society},
  doi = {10.1103/PhysRevB.110.L220409},
  url = {https://link.aps.org/doi/10.1103/PhysRevB.110.L220409}
}

@article{Sun2024Dipolar,
  author       = {Yue Sun and Fanhao Meng and Changmin Lee and Aljoscha Soll and Hongrui Zhang and Ramamoorthy Ramesh and Jie Yao and Zden{\v e}k Sofer and Joseph Orenstein},
  title        = {Dipolar spin wave packet transport in a van der Waals antiferromagnet},
  journal      = {Nature Physics},
  volume       = {20},
  number       = {5},
  pages        = {794--800},
  year         = {2024},
  doi          = {10.1038/s41567-024-02387-2},
}

@article{Qin2023RoomTemperature,
  author       = {Peixin Qin and Han Yan and Xiaoning Wang and Hongyu Chen and Ziang Meng and
                  Jianting Dong and Meng Zhu and Jialin Cai and Zexin Feng and Xiaorong Zhou and
                  Li Liu and Tianli Zhang and Zhongming Zeng and Jia Zhang and Chengbao Jiang and Zhiqi Liu},
  title        = {Room-temperature magnetoresistance in an all-antiferromagnetic tunnel junction},
  journal      = {Nature},
  volume       = {613},
  number       = {7944},
  pages        = {485--489},
  year         = {2023},
  doi          = {10.1038/s41586-022-05461-y},
}

@article{Chen2023OctupoleDriven,
  author       = {Xianzhe Chen and Tomoya Higo and Katsuhiro Tanaka and Takuya Nomoto and 
                  Hanshen Tsai and Hiroshi Idzuchi and Masanobu Shiga and Shoya Sakamoto and 
                  Ryoya Ando and Hidetoshi Kosaki and Takumi Matsuo and Daisuke Nishio-Hamane and 
                  Ryotaro Arita and Shinji Miwa and Satoru Nakatsuji},
  title        = {Octupole-driven magnetoresistance in an antiferromagnetic tunnel junction},
  journal      = {Nature},
  volume       = {613},
  number       = {7944},
  pages        = {490--495},
  year         = {2023},
  doi          = {10.1038/s41586-022-05463-w},
}

@article{PhysRevApplied.21.054002,
  title = {Double-free-layer stochastic magnetic tunnel junctions with synthetic antiferromagnets},
  author = {Selcuk, Kemal and Kanai, Shun and Ota, Rikuto and Ohno, Hideo and Fukami, Shunsuke and Camsari, Kerem Y.},
  journal = {Phys. Rev. Appl.},
  volume = {21},
  issue = {5},
  pages = {054002},
  numpages = {12},
  year = {2024},
  month = {May},
  publisher = {American Physical Society},
  doi = {10.1103/PhysRevApplied.21.054002},
  url = {https://link.aps.org/doi/10.1103/PhysRevApplied.21.054002}
}

@article{Nemec2018AntiferroOptospintronics,
  author       = {Petr Němec and Mathias Fiebig and Thomas Kampfrath and Alexey V. Kimel},
  title        = {Antiferromagnetic opto‐spintronics},
  journal      = {Nature Physics},
  volume       = {14},
  number       = {3},
  pages        = {229--241},
  year         = {2018},
  doi          = {10.1038/s41567-018-0051-x},
}

@article{Sierra2021vdWSpintronics,
  author       = {Juan F. Sierra and Jaroslav Fabian and Roland K. Kawakami and Stephan Roche and Sergio O. Valenzuela},
  title        = {Van der Waals heterostructures for spintronics and opto‑spintronics},
  journal      = {Nature Nanotechnology},
  volume       = {16},
  number       = {8},
  pages        = {856--868},
  year         = {2021},
  doi          = {10.1038/s41565-021-00936-x},
}

@article{PhysRevLett.116.097204,
  title = {Antiferromagnetic Spin Seebeck Effect},
  author = {Wu, Stephen M. and Zhang, Wei and KC, Amit and Borisov, Pavel and Pearson, John E. and Jiang, J. Samuel and Lederman, David and Hoffmann, Axel and Bhattacharya, Anand},
  journal = {Phys. Rev. Lett.},
  volume = {116},
  issue = {9},
  pages = {097204},
  numpages = {5},
  year = {2016},
  month = {Mar},
  publisher = {American Physical Society},
  doi = {10.1103/PhysRevLett.116.097204},
  url = {https://link.aps.org/doi/10.1103/PhysRevLett.116.097204}
}

@book{Dey2021Spintronics,
  author       = {Puja Dey and Jitendra Nath Roy},
  title        = {Spintronics: Fundamentals and Applications},
  publisher    = {Springer Singapore},
  year         = {2021},
  edition      = {1},
  isbn         = {978-981-16-0069-2},
  doi          = {10.1007/978-981-16-0069-2},
  pages        = {XX, 273},
  note         = {Published: 13 April 2021},
}

@article{Wang2021SpinPL,
  author       = {Wang, X. and Cao, J. and Lu, Z. and Cohen, A. and Kitadai, H. and Li, T. and 
                  Tan, Q. and Wilson, M. and Lui, C. H. and Smirnov, D. and Sharifzadeh, S. and 
                  Ling, X.},
  title        = {Spin-Induced Linear Polarization of Photoluminescence in Antiferromagnetic van der Waals Crystals},
  journal      = {Nature Materials},
  year         = {2021},
  volume       = {20},
  number       = {7},
  pages        = {964--970},
  doi          = {10.1038/s41563-021-00968-7}
}

@article{Kang2020NiPS3,
  author       = {Kang, S. and Kim, K. and Kim, B. H. and Kim, J. and Sim, K. I. and Lee, J.-U. and 
                  Lee, S. and Park, K. and Yun, S. and Kim, T. and Nag, A. and Walters, A. and 
                  Garcia-Fernandez, M. and Li, J. and Chapon, L. and Zhou, K.-J. and Son, Y.-W. and 
                  Kim, J. H. and Cheong, H. and Park, J.-G.},
  title        = {Coherent Many-Body Exciton in van der Waals Antiferromagnet NiPS\textsubscript{3}},
  journal      = {Nature},
  year         = {2020},
  volume       = {583},
  number       = {7818},
  pages        = {785--789},
  doi          = {10.1038/s41586-020-2520-5}
}

@article{PhysRevB.104.054419,
  title = {Magnon-driven skyrmion dynamics in antiferromagnets: Effect of magnon polarization},
  author = {Jin, Z. and Meng, C. Y. and Liu, T. T. and Chen, D. Y. and Fan, Z. and Zeng, M. and Lu, X. B. and Gao, X. S. and Qin, M. H. and Liu, J.-M.},
  journal = {Phys. Rev. B},
  volume = {104},
  issue = {5},
  pages = {054419},
  numpages = {8},
  year = {2021},
  month = {Aug},
  publisher = {American Physical Society},
  doi = {10.1103/PhysRevB.104.054419},
  url = {https://link.aps.org/doi/10.1103/PhysRevB.104.054419}
}

@article{Telford2020CrSBr,
  author       = {Telford, E. J. and Dismukes, A. H. and Lee, K. and Cheng, M. and Wieteska, A. and 
                  Bartholomew, A. K. and Chen, Y.-S. and Xu, X. and Pasupathy, A. N. and Zhu, X. and 
                  Dean, C. R. and Roy, X.},
  title        = {Layered Antiferromagnetism Induces Large Negative Magnetoresistance in the van der Waals Semiconductor CrSBr},
  journal      = {Advanced Materials},
  year         = {2020},
  volume       = {32},
  number       = {37},
  pages        = {2003240},
  doi          = {10.1002/adma.202003240}
}

@article{Lee2021CrSBr,
  author       = {Lee, K. and Dismukes, A. H. and Telford, E. J. and Wiscons, R. A. and Wang, J. and 
                  Xu, X. and Nuckolls, C. and Dean, C. R. and Roy, X. and Zhu, X.},
  title        = {Magnetic Order and Symmetry in the 2D Semiconductor CrSBr},
  journal      = {Nano Letters},
  year         = {2021},
  volume       = {21},
  number       = {8},
  pages        = {3511--3517},
  doi          = {10.1021/acs.nanolett.1c00219}
}

@article{Wilson2021Interlayer,
  author       = {Wilson, N. P. and Lee, K. and Cenker, J. and Xie, K. and Dismukes, A. H. and 
                  Telford, E. J. and Fonseca, J. and Sivakumar, S. and Dean, C. and Cao, T. and 
                  Roy, X. and Xu, X. and Zhu, X.},
  title        = {Interlayer Electronic Coupling on Demand in a 2D Magnetic Semiconductor},
  journal      = {Nature Materials},
  year         = {2021},
  volume       = {20},
  number       = {12},
  pages        = {1657--1662},
  doi          = {10.1038/s41563-021-01070-8}
}

@article{He2024AFMSkyrmionNanoLett,
  author       = {He, Bin and Jin, Haonan and Zheng, Dongfeng and Liu, Yizhou and Li, Jialiang and Hu, Yue and 
                  Wang, Yuqiang and Zhang, Junwei and Peng, Yong and Wan, Caihua and Zhu, Tao and Han, Xiufeng and 
                  Zhang, Shilei and Yu, Guoqiang},
  title        = {Creation of Room‑Temperature Sub‑100 nm Antiferromagnetic Skyrmions in an Antiferromagnet IrMn Through Interfacial Exchange Coupling},
  journal      = {Nano Letters},
  year         = {2024},
  volume       = {24},
  number       = {7},
  pages        = {2196--2202},
  doi          = {10.1021/acs.nanolett.3c04221}
}

@article{Olsen2024Antiferromagnetism2D,
  author       = {Olsen, Thomas},
  title        = {Antiferromagnetism in two-dimensional materials: progress and computational challenges},
  journal      = {2D Materials},
  year         = {2024},
  volume       = {11},
  number       = {3},
  pages        = {033005},
  doi          = {10.1088/2053-1583/ad4ef1},
  note         = {Published online: 14 June 2024},
}

@article{PhysRevB.90.085429,
  title = {Graphene spintronics: Spin injection and proximity effects from first principles},
  author = {Lazi\ifmmode \acute{c}\else \'{c}\fi{}, P. and Sipahi, G. M. and Kawakami, R. K. and \ifmmode \check{Z}\else \v{Z}\fi{}uti\ifmmode \acute{c}\else \'{c}\fi{}, Igor},
  journal = {Phys. Rev. B},
  volume = {90},
  issue = {8},
  pages = {085429},
  numpages = {15},
  year = {2014},
  month = {Aug},
  publisher = {American Physical Society},
  doi = {10.1103/PhysRevB.90.085429},
  url = {https://link.aps.org/doi/10.1103/PhysRevB.90.085429}
}

@article{Nair2013DefectMagnetismGraphene,
  author       = {R. R. Nair and I.-L. Tsai and M. Sepioni and O. Lehtinen and J. Keinonen and A. V. Krasheninnikov and A. H. Castro Neto and M. I. Katsnelson and A. K. Geim and I. V. Grigorieva},
  title        = {Dual origin of defect magnetism in graphene and its reversible switching by molecular doping},
  journal      = {Nature Communications},
  volume       = {4},
  article      = {2010},
  year         = {2013},
  doi          = {10.1038/ncomms3010},
  url          = {https://www.nature.com/articles/ncomms3010}
}

@article{Kohn1965SelfConsistent,
  author       = {Walter Kohn and Lu Jeu Sham},
  title        = {Self-Consistent Equations Including Exchange and Correlation Effects},
  journal      = {Physical Review},
  volume       = {140},
  number       = {4A},
  pages        = {A1133--A1138},
  year         = {1965},
  doi          = {10.1103/PhysRev.140.A1133},
}

@article{Dederichs1984GroundStates,
  author       = {P. H. Dederichs and S. Blügel and R. Zeller and H. Akai},
  title        = {Ground States of Constrained Systems: Application to Cerium Impurities},
  journal      = {Physical Review Letters},
  volume       = {53},
  number       = {26},
  pages        = {2512--2515},
  year         = {1984},
  doi          = {10.1103/PhysRevLett.53.2512},
}

@article{Liechtenstein1987ExchangeInteractions,
  author       = {A. I. Liechtenstein and M. I. Katsnelson and V. P. Antropov and V. A. Gubanov},
  title        = {Local Spin Density Functional Approach to the Theory of Exchange Interactions in Ferromagnetic Metals and Alloys},
  journal      = {Journal of Magnetism and Magnetic Materials},
  volume       = {67},
  number       = {1},
  pages        = {65--74},
  year         = {1987},
  doi          = {10.1016/0304-8853(87)90696-7},
}

@article{Sverdlov2015SiliconSpintronics,
  author       = {Viktor Sverdlov and Siegfried Selberherr},
  title        = {Silicon spintronics: Progress and challenges},
  journal      = {Physics Reports},
  volume       = {585},
  pages        = {1--40},
  year         = {2015},
  issn         = {0370-1573},
  doi          = {10.1016/j.physrep.2015.05.002},
  url          = {https://www.sciencedirect.com/science/article/pii/S0370157315002409}
}

@article{Wang2021OhmicContacts,
  author       = {Qing Wang and et al.},
  title        = {Efficient Ohmic contacts and built-in atomic sublayer protection in MoSi$_2$N$_4$ and WSi$_2$N$_4$ monolayers},
  journal      = {npj 2D Materials and Applications},
  volume       = {5},
  number       = {1},
  year         = {2021},
  month        = {December},
  doi          = {10.1038/s41699-021-00251-y},
}

@article{Ai2021SpinValleyMoSi2X4,
  author       = {Heng Ai and Dong Liu and Jie Geng and Sheng Wang and K. H. Lo and Hong Pan},
  title        = {Theoretical Evidence of the Spin-Valley Coupling and Valley Polarization in Two-Dimensional MoSi$_2$X$_4$ (X = N, P, and As)},
  journal      = {Physical Chemistry Chemical Physics},
  volume       = {23},
  number       = {4},
  pages        = {3144--3151},
  year         = {2021},
  doi          = {10.1039/d0cp05926a},
}

@article{Dey2022IntrinsicFM,
  author       = {D. Dey and A. Ray and L. Yu},
  title        = {Intrinsic Ferromagnetism and Restrictive Thermodynamic Stability in MA$_2$N$_4$ and Janus VSiGeN$_4$ Monolayers},
  journal      = {Physical Review Materials},
  volume       = {6},
  number       = {6},
  year         = {2022},
  doi          = {10.1103/PhysRevMaterials.6.L061002},
}

@article{Zhang2022JanusMA2Z4,
  author       = {W. Zhang and W. Yang and Y. Liu and Z. Liu and F. Zhang},
  title        = {Computational Exploration and Screening of Novel Janus MA$_2$Z$_4$ (M = Sc–Zn, Y–Ag, Hf–Au; A = Si, Ge; Z = N, P) Monolayers and Potential Application as a Photocatalyst},
  journal      = {Frontiers of Physics},
  volume       = {17},
  number       = {6},
  year         = {2022},
  doi          = {10.1007/s11467-022-1199-5},
}

@article{Liu2022StrainMoSi2N4,
  author       = {C. Liu and Z. Wang and W. Xiong and H. Zhong and S. Yuan},
  title        = {Effect of Vertical Strain and In-Plane Biaxial Strain on Type-II MoSi$_2$N$_4$/Cs$_3$Bi$_2$I$_9$ van der Waals Heterostructure},
  journal      = {Journal of Applied Physics},
  volume       = {131},
  number       = {16},
  year         = {2022},
  doi          = {10.1063/5.0080224},
}

@article{Shu2022PhotogalvanicDetector,
  author       = {L. Shu and L. Qian and X. Ye and Y. Xie},
  title        = {Multifunctional Two-Dimensional VSi$_2$N$_4$/WSi$_2$N$_4$/VSi$_2$N$_4$ Photodetector Driven by the Photogalvanic Effect},
  journal      = {Physical Review Applied},
  volume       = {17},
  number       = {5},
  pages        = {054010},
  year         = {2022},
  doi          = {10.1103/PhysRevApplied.17.054010},
}

@article{Su2023MoSi2N4Nanoribbons,
  author       = {X. Q. Su and X. F. Wang},
  title        = {Electronic and Spintronic Properties of Armchair MoSi$_2$N$_4$ Nanoribbons Doped by 3D Transition Metals},
  journal      = {Nanomaterials},
  volume       = {13},
  number       = {4},
  pages        = {676},
  year         = {2023},
  doi          = {10.3390/nano13040676},
}

@article{Zhang2024SpinFET,
  author       = {X. Zhang and B. Liu and J. Huang and X. Cao and Y. Zhang and Z. X. Guo},
  title        = {Nonvolatile Spin Field Effect Transistor Based on VSi$_2$N$_4$/Sc$_2$CO$_2$ Multiferroic Heterostructure},
  journal      = {Physical Review B},
  volume       = {109},
  number       = {20},
  pages        = {205105},
  year         = {2024},
  doi          = {10.1103/PhysRevB.109.205105},
}

@article{Ahmad2024TiSi2N4,
  author       = {S. Ahmad and H. U. Din and C. Q. Nguyen and S. T. Nguyen and C. Nguyen},
  title        = {Alkali to alkaline earth metals: a DFT study of monolayer TiSi$_2$N$_4$ for metal ion batteries},
  journal      = {Dalton Transactions},
  volume       = {53},
  number       = {8},
  pages        = {3785--3796},
  year         = {2024},
  doi          = {10.1039/d3dt03946c},
}

@article{Yang2021ValleyPseudospin,
  author       = {C. Yang and Z. Song and X. Sun and J. Lu},
  title        = {Valley pseudospin in monolayer MoSi$_2$N$_4$ and MoSi$_2$As$_4$},
  journal      = {Physical Review B},
  volume       = {103},
  number       = {3},
  pages        = {035308},
  year         = {2021},
  doi          = {10.1103/PhysRevB.103.035308},
}

@article{Liang2023ValleySpinSplittings,
  author       = {L. Liang and Y. Yang and X. Wang and X. Li},
  title        = {Tunable Valley and Spin Splittings in VSi$_2$N$_4$ Bilayers},
  journal      = {Nano Letters},
  volume       = {23},
  number       = {3},
  pages        = {858--862},
  year         = {2023},
  doi          = {10.1021/acs.nanolett.2c03963},
}

@article{Wang2021MA2Z4,
  author       = {L. Wang and et al.},
  title        = {Intercalated architecture of MA$_2$Z$_4$ family layered van der Waals materials with emerging topological, magnetic and superconducting properties},
  journal      = {Nature Communications},
  volume       = {12},
  number       = {1},
  pages        = {4--13},
  year         = {2021},
  doi          = {10.1038/s41467-021-22324-8},
}

@article{Wang2022VSi2N4Anode,
  author       = {Z. Wang and et al.},
  title        = {Heavy 2D VSi$_2$N$_4$: High Capacity and Full Battery Open-Circuit Voltage as Li/Na-Ion Batteries Anode},
  journal      = {Applied Surface Science},
  volume       = {593},
  year         = {2022},
  pages        = {153354},
  doi          = {10.1016/j.apsusc.2022.153354},
}

@article{Wu2022GiantTMR,
  author       = {Q. Wu and L. K. Ang},
  title        = {Giant tunneling magnetoresistance in atomically thin VSi$_2$N$_4$/MoSi$_2$N$_4$/VSi$_2$N$_4$ magnetic tunnel junction},
  journal      = {Applied Physics Letters},
  volume       = {120},
  number       = {2},
  year         = {2022},
  doi          = {10.1063/5.0075046},
}

@article{Zhang2023Ti3C2T2TiSi2N4,
  author       = {Z. Zhang and J. Wang and Z. Dai and M. Zhang and L. Niu and L. Bai},
  title        = {Modulation of Contact Types and Schottky Barrier in Ti$_3$C$_2$T$_2$/TiSi$_2$N$_4$ (T = O or OH) van der Waals Heterostructures by Biaxial Strain and External Electric Field},
  journal      = {Chemical Physics},
  volume       = {573},
  year         = {2023},
  pages        = {111996},
  doi          = {10.1016/j.chemphys.2023.111996},
}

@article{Tian2022ElectronicPropertiesMnSi2N4,
  author       = {M. Tian and C. Wei and J. Zhang and Z. Wang},
  title        = {Electronic Properties and Storage Capability of Two-Dimensional Nitridosilicate MnSi$_2$N$_4$ from First-Principles},
  journal      = {AIP Advances},
  year         = {2022},
  doi          = {10.1063/5.0127013},
}

@article{PhysRevLett.17.1133,
  title = {Absence of Ferromagnetism or Antiferromagnetism in One- or Two-Dimensional Isotropic Heisenberg Models},
  author = {Mermin, N. D. and Wagner, H.},
  journal = {Phys. Rev. Lett.},
  volume = {17},
  issue = {22},
  pages = {1133--1136},
  numpages = {0},
  year = {1966},
  month = {Nov},
  publisher = {American Physical Society},
  doi = {10.1103/PhysRevLett.17.1133},
  url = {https://link.aps.org/doi/10.1103/PhysRevLett.17.1133}
}

@article{Wei2025UpperCriticalFields,
  author       = {Wei Wei and Yuling Xiang and Qiang Hou and Yue Sun and Zhixiang Shi},
  title        = {Upper critical fields in high-temperature superconductors},
  journal      = {Journal of Physics: Condensed Matter},
  volume       = {37},
  number       = {14},
  pages        = {143003},
  year         = {2025},
  doi          = {10.1088/1361-648X/adb2d2},
  publisher    = {IOP Publishing Ltd},
  note         = {Topical Review},
}

@article{PhysRevB.111.155122,
  author       = {H. Park and G. Kotliar and H. Lee},
  title        = {Role of magnetic and structural symmetry breaking in forming the Mott insulating gap},
  journal      = {Physical Review B},
  volume       = {111},
  issue        = {15},
  pages        = {155122},
  year         = {2025},
  doi          = {10.1103/PhysRevB.111.155122},
}

@article{Chou2024SpinPolarization,
  author    = {Chung-Tao Chou and Supriya Ghosh and Brooke C. McGoldrick and Thanh Nguyen and Gautam Gurung and Evgeny Y. Tsymbal and Mingda Li and K. Andre Mkhoyan and Luqiao Liu},
  title     = {Large Spin Polarization from symmetry-breaking Antiferromagnets in Antiferromagnetic Tunnel Junctions},
  journal   = {Nature Communications},
  volume    = {15},
  article   = {7840},
  year      = {2024},
  doi       = {10.1038/s41467-024-52208-6},
}

@article{Smejkal2020CrystalTRSB,
  author       = {Libor Šmejkal and Tomáš Jungwirth and Jairo Sinova and et al.},
  title        = {Crystal time-reversal symmetry breaking and spontaneous Hall effect in collinear antiferromagnets},
  journal      = {Science Advances},
  volume       = {6},
  number       = {39},
  pages        = {eaaz8809},
  year         = {2020},
  doi          = {10.1126/sciadv.aaz8809},
}

@article{Ren2023MX2Y4,
  author    = {K. Ren and H. Shu and K. Wang and H. Qin},
  title     = {Two-dimensional MX$_2$Y$_4$ systems: ultrahigh carrier transport and excellent hydrogen evolution reaction performances},
  journal   = {Physical Chemistry Chemical Physics},
  volume    = {25},
  number    = {6},
  pages     = {4519--4527},
  year      = {2023},
  doi       = {10.1039/d2cp04224j}
}

@article{PedrozaRojas2025,
  author    = {Pedroza-Rojas, Brandon and Sanchez-Castillo, Ariadna and Ponce-P\'erez, Rodrigo},
  title     = {Structural, Electronic, and Magnetic Properties of the van der Waals ScSi$_2$N$_4$/VSi$_2$N$_4$ Heterostructure: A First-Principles Study},
  journal   = {ACS Omega},
  year      = {2025},
  volume    = {10},
  number    = {21},
  pages     = {22062--22070},
  doi       = {10.1021/acsomega.5c02195},
  url       = {https://pubs.acs.org/doi/10.1021/acsomega.5c02195},
  note      = {Open Access, Published May 20, 2025}
}

@article{BernalWalker1997,
  author    = {J. D. Bernal and S. J. Walker},
  title     = {The Structure of Graphite},
  journal   = {Proceedings of the Royal Society of London. Series A, Containing Papers of a Mathematical and Physical Character},
  volume    = {1},
  number    = {1855},
  pages     = {1123--1125},
  year      = {1997}
}

@article{VanDerZande2013,
  author    = {A. M. Van Der Zande and others},
  title     = {Grains and grain boundaries in highly crystalline monolayer molybdenum disulphide},
  journal   = {Nature Materials},
  volume    = {12},
  number    = {6},
  pages     = {554--561},
  year      = {2013},
  doi       = {10.1038/nmat3633}
}

@article{Zutic2019Proximitized,
  author       = {Žutić, Igor and Matos-Abiague, Alex and Scharf, Benedikt and Dery, Hanan and Belashchenko, Kirill},
  title        = {Proximitized materials},
  journal      = {Materials Today},
  volume       = {22},
  number       = {February},
  pages        = {85--107},
  year         = {2019},
  doi          = {10.1016/j.mattod.2018.05.003}
}

@article{Zhao2023,
  author    = {Z. Zhao and X. Duan and X. Fang and X. Wang and W. Mi},
  title     = {Prediction of electronic structure and magnetic anisotropy of two-dimensional M Si$_2$N$_4$ (M = 3d transition-metal) monolayers},
  journal   = {Applied Surface Science},
  volume    = {611},
  pages     = {155693},
  year      = {2023},
  doi       = {10.1016/j.apsusc.2022.155693}
}

@article{PhysRevB.105.014437,
  title = {Large magneto-optical effect and magnetic anisotropy energy in two-dimensional metallic ferromagnet ${\mathrm{Fe}}_{3}\mathrm{Ge}{\mathrm{Te}}_{2}$},
  author = {Jiang, Ming-Chun and Guo, Guang-Yu},
  journal = {Phys. Rev. B},
  volume = {105},
  issue = {1},
  pages = {014437},
  numpages = {14},
  year = {2022},
  month = {Jan},
  publisher = {American Physical Society},
  doi = {10.1103/PhysRevB.105.014437},
  url = {https://link.aps.org/doi/10.1103/PhysRevB.105.014437}
}

@article{hnm7-xz31,
  title = {Large magnetic anisotropy and high spin state in two-dimensional ${\mathrm{MnIn}}_{2}{\mathrm{Se}}_{3}{\mathrm{I}}_{2}$},
  author = {Yang, Yanyan and Chen, Xinyu and Zhang, Guitao and Li, Kang and Wang, Jinlan and Chen, Qian},
  journal = {Phys. Rev. B},
  volume = {112},
  issue = {18},
  pages = {184426},
  numpages = {8},
  year = {2025},
  month = {Nov},
  publisher = {American Physical Society},
  doi = {10.1103/hnm7-xz31},
  url = {https://link.aps.org/doi/10.1103/hnm7-xz31}
}

@article{Chen2022,
  author    = {D. Chen and others},
  title     = {Electrical and magnetic properties of antiferromagnetic semiconductor MnSi$_2$N$_4$ monolayer},
  journal   = {Frontiers in Chemistry},
  year      = {2022},
  pages     = {1--7},
  doi       = {10.3389/fchem.2022.1103704}
}

@article{An2023,
  author    = {Z. An and L. Lv and Y. Su and Y. Jiang and Z. Guan},
  title     = {Carrier doping modulates the magnetoelectronic and magnetic anisotropic properties of two-dimensional MSi$_2$N$_4$ (M = Cr, Mn, Fe, and Co) monolayers},
  journal   = {Physical Chemistry Chemical Physics},
  volume    = {26},
  number    = {5},
  pages     = {4208--4217},
  year      = {2023},
  doi       = {10.1039/d3cp05032g}
}

@article{Tung2016,
  author  = {R. T. Tung and L. Kronik},
  title   = {Band offset formation at semiconductor heterojunctions through density-based minimization of interface energy},
  journal = {Physical Review B},
  volume  = {94},
  number  = {7},
  pages   = {1--24},
  year    = {2016},
  doi     = {10.1103/PhysRevB.94.075310}
}

@article{giannozzi2009,
  author  = {P. Giannozzi and others},
  title   = {QUANTUM ESPRESSO: A modular and open-source software project for quantum simulations of materials},
  journal = {J. Phys.: Condens. Matter},
  volume  = {21},
  number  = {39},
  year    = {2009},
  doi     = {10.1088/0953-8984/21/39/395502}
}

@article{giannozzi2020,
  author  = {P. Giannozzi and others},
  title   = {Quantum ESPRESSO toward the exascale},
  journal = {J. Chem. Phys.},
  volume  = {152},
  number  = {15},
  year    = {2020},
  doi     = {10.1063/5.0005082}
}

@article{blochl1994,
  author  = {P. E. Blöchl},
  title   = {Projector augmented-wave method},
  journal = {Phys. Rev. B},
  volume  = {50},
  number  = {24},
  year    = {1994},
  pages   = {17953},
  doi     = {10.1103/PhysRevB.50.17953}
}

@article{joubert1999,
  author  = {D. Joubert},
  title   = {From ultrasoft pseudopotentials to the projector augmented-wave method},
  journal = {Phys. Rev. B},
  volume  = {59},
  number  = {3},
  year    = {1999},
  pages   = {1758--1775},
  doi     = {10.1103/PhysRevB.59.1758}
}

@article{perdew1996,
  author  = {J. P. Perdew and K. Burke and M. Ernzerhof},
  title   = {Generalized Gradient Approximation Made Simple},
  journal = {Phys. Rev. Lett.},
  volume  = {77},
  number  = {18},
  year    = {1996},
  pages   = {3865--3868},
  doi     = {10.1103/PhysRevLett.77.3865}
}

@article{grimme2006,
  author  = {S. Grimme},
  title   = {Semiempirical GGA-Type Density Functional Constructed with a Long-Range Dispersion Correction},
  journal = {J. Comput. Chem.},
  volume  = {27},
  year    = {2006},
  pages   = {1787--1799},
  doi     = {10.1002/jcc.20495}
}

@article{grimme2010,
  author  = {S. Grimme and J. Antony and S. Ehrlich and H. Krieg},
  title   = {A consistent and accurate ab initio parametrization of density functional dispersion correction (DFT-D) for the 94 elements H–Pu},
  journal = {J. Chem. Phys.},
  volume  = {132},
  number  = {15},
  year    = {2010},
  doi     = {10.1063/1.3382344}
}

@article{grimme2011,
  author  = {S. Grimme and S. Ehrlich and L. Goerigk},
  title   = {Effect of the damping function in dispersion corrected density functional theory},
  journal = {J. Comput. Chem.},
  volume  = {32},
  year    = {2011},
  pages   = {1456--1465},
  doi     = {10.1002/jcc.21759}
}

@article{becke2005,
    author = {Becke, Axel D. and Johnson, Erin R.},
    title = {A density-functional model of the dispersion interaction},
    journal = {The Journal of Chemical Physics},
    volume = {123},
    number = {15},
    pages = {154101},
    year = {2005},
    month = {10},
    issn = {0021-9606},
    doi = {10.1063/1.2065267},
    url = {https://doi.org/10.1063/1.2065267},
    eprint = {https://pubs.aip.org/aip/jcp/article-pdf/doi/10.1063/1.2065267/13408810/154101_1_online.pdf},
}

@article{hubbard1963,
  author  = {J. Hubbard},
  title   = {Electron Correlations in Narrow Energy Bands},
  journal = {Proc. R. Soc. Lond. A},
  volume  = {276},
  number  = {1365},
  year    = {1963},
  pages   = {238--257},
  doi     = {10.1098/rspa.1963.0204}
}

@article{aikebaier2015,
  author  = {F. Aikebaier and A. Pertsova and C. M. Canali},
  title   = {Effects of short-range electron-electron interactions in doped graphene},
  journal = {Phys. Rev. B},
  volume  = {92},
  number  = {15},
  year    = {2015},
  doi     = {10.1103/PhysRevB.92.155420}
}

@article{monkhorst1976,
  author  = {H. J. Monkhorst and J. D. Pack},
  title   = {Special points for Brillouin-zone integrations},
  journal = {Phys. Rev. B},
  volume  = {13},
  number  = {12},
  year    = {1976},
  pages   = {5188--5192},
  doi     = {10.1103/PhysRevB.13.5188}
}

@article{https://doi.org/10.1002/adma.202310379,
author = {Chen, Hongyu and Liu, Li and Zhou, Xiaorong and Meng, Ziang and Wang, Xiaoning and Duan, Zhiyuan and Zhao, Guojian and Yan, Han and Qin, Peixin and Liu, Zhiqi},
title = {Emerging Antiferromagnets for Spintronics},
journal = {Advanced Materials},
volume = {36},
number = {14},
pages = {2310379},
doi = {https://doi.org/10.1002/adma.202310379},
url = {https://advanced.onlinelibrary.wiley.com/doi/abs/10.1002/adma.202310379},
eprint = {https://advanced.onlinelibrary.wiley.com/doi/pdf/10.1002/adma.202310379},
year = {2024}
}

@article{facemyer2023,
  title = {{Spin and electronic excitations in $4f$ atomic chains on Au(111) substrates}},
  author = {Facemyer, David W. and Dandu, Naveen K. and Lee, Alex Taekyung and Singh, Vijay R. and Ngo, Anh T. and Ulloa, Sergio E.},
  journal = {Phys. Rev. B},
  volume = {108},
  issue = {8},
  pages = {085407},
  numpages = {8},
  year = {2023},
  month = {Aug},
  publisher = {American Physical Society},
  doi = {10.1103/PhysRevB.108.085407},
  url = {https://link.aps.org/doi/10.1103/PhysRevB.108.085407}
}

@article{facemyer2025,
  title = {Probing nonlocal correlations in magnetic rare-earth clusters},
  author = {Facemyer, David W. and Ulloa, Sergio E.},
  journal = {Phys. Rev. B},
  volume = {111},
  issue = {6},
  pages = {064403},
  numpages = {8},
  year = {2025},
  month = {Feb},
  publisher = {American Physical Society},
  doi = {10.1103/PhysRevB.111.064403},
  url = {https://link.aps.org/doi/10.1103/PhysRevB.111.064403}
}

@article{choi2019,
  title = {Colloquium: Atomic spin chains on surfaces},
  author = {Choi, Deung-Jang and Lorente, Nicolas and Wiebe, Jens and von Bergmann, Kirsten and Otte, Alexander F. and Heinrich, Andreas J.},
  journal = {Rev. Mod. Phys.},
  volume = {91},
  issue = {4},
  pages = {041001},
  numpages = {24},
  year = {2019},
  month = {Oct},
  publisher = {American Physical Society},
  doi = {10.1103/RevModPhys.91.041001},
  url = {https://link.aps.org/doi/10.1103/RevModPhys.91.041001}}

@article{Zhou2025AFMtoAM,
  author       = {Zhou, P. and Peng, X. N. and Hu, Y. Z. and Pan, B. R. and Liu, S. M. and Lyu, Pengbo and Sun, L. Z.},
  title        = {Transition from antiferromagnets to altermagnets: Symmetry-Breaking Theory},
  journal      = {Physical Review B},
  volume       = {112},
  pages        = {144419},
  year         = {2025},
  doi          = {10.1103/PhysRevB.112.144419},
  eprint       = {arXiv:2410.17747},
  archivePrefix= {arXiv},
  primaryClass = {cond-mat.mtrl-sci},
  note         = {Last revised 15 Oct 2025 (v3)},
  url          = {https://arxiv.org/abs/2410.17747}
}

@article{Xiang2013,
  title = {Magnetic properties and energy-mapping analysis},
  author = {Xiang, H. J. and Lee, C. and Koo, H.-J. and Gong, X. G. and Whangbo, M.-H.},
  journal = {Dalton Trans.},
  volume = {42},
  number = {4},
  pages = {823--853},
  year = {2013},
  publisher = {Royal Society of Chemistry},
  doi = {10.1039/C2DT31662E}
}

@article{
doi:10.1126/sciadv.aaz8809,
author = {Libor Šmejkal  and Rafael González-Hernández  and T. Jungwirth  and J. Sinova },
title = {Crystal time-reversal symmetry breaking and spontaneous Hall effect in collinear antiferromagnets},
journal = {Science Advances},
volume = {6},
number = {23},
pages = {eaaz8809},
year = {2020},
doi = {10.1126/sciadv.aaz8809},
URL = {https://www.science.org/doi/abs/10.1126/sciadv.aaz8809},
eprint = {https://www.science.org/doi/pdf/10.1126/sciadv.aaz8809}}

@article{hiendrich,
	title = {Nanoscale mechanics of antiferromagnetic domain walls},
	volume = {17},
	issn = {1745-2481},
	url = {https://doi.org/10.1038/s41567-020-01157-0},
	doi = {10.1038/s41567-020-01157-0},
	number = {5},
	journal = {Nature Physics},
	author = {Hedrich, Natascha and Wagner, Kai and Pylypovskyi, Oleksandr V. and Shields, Brendan J. and Kosub, Tobias and Sheka, Denis D. and Makarov, Denys and Maletinsky, Patrick},
	month = may,
	year = {2021},
	pages = {574--577},
}

@article{Zur2023,
  title = {Magnetic Imaging and Domain Nucleation in {CrSBr} Down to the {2D} Limit},
  author = {Zur, Yishay and Noah, Avia and Boix-Constant, Carla and Ma{\~n}as-Valero, Samuel and Fridman, Nofar and Rama-Eiroa, Ricardo and Huber, Martin E. and Santos, Elton J. G. and Coronado, Eugenio and Anahory, Yonathan},
  journal = {Advanced Materials},
  volume = {36},
  number = {2},
  pages = {2307195},
  year = {2023},
  doi = {10.1002/adma.202307195},
  url = {https://doi.org/10.1002/adma.202307195}
}

@article{tschudin_imaging_2024,
	title = {Imaging nanomagnetism and magnetic phase transitions in atomically thin {CrSBr}},
	volume = {15},
	issn = {2041-1723},
	url = {https://doi.org/10.1038/s41467-024-49717-9},
	doi = {10.1038/s41467-024-49717-9},
	number = {1},
	journal = {Nature Communications},
	author = {Tschudin, Märta A. and Broadway, David A. and Siegwolf, Patrick and Schrader, Carolin and Telford, Evan J. and Gross, Boris and Cox, Jordan and Dubois, Adrien E. E. and Chica, Daniel G. and Rama-Eiroa, Ricardo and J. G. Santos, Elton and Poggio, Martino and Ziebel, Michael E. and Dean, Cory R. and Roy, Xavier and Maletinsky, Patrick},
	month = jul,
	year = {2024},
	pages = {6005},
}

@article{polymorphism_CSB,
	title = {Resolving and routing magnetic polymorphs in a {2D} layered antiferromagnet},
	volume = {24},
	issn = {1476-4660},
	url = {https://doi.org/10.1038/s41563-024-02074-w},
	doi = {10.1038/s41563-024-02074-w},
	number = {2},
	journal = {Nature Materials},
	author = {Sun, Zeyuan and Hong, Canyu and Chen, Yi and Sheng, Zhiyuan and Wu, Shuang and Wang, Zhanshan and Liang, Bokai and Liu, Wei-Tao and Yuan, Zhe and Wu, Yizheng and Mi, Qixi and Liu, Zhongkai and Shen, Jian and Wu, Shiwei},
	month = feb,
	year = {2025},
	pages = {226--233},
}

@article{Chen2024,
  title = {Even--Odd Layer-Dependent Exchange Bias Effect in {MnBi$_2$Te$_4$} Chern Insulator Devices},
  author = {Chen, Bo and Liu, Xiaoda and Li, Yuhang and Tay, Han and Taniguchi, Takashi and Watanabe, Kenji and Chan, Moses H. W. and Yan, Jiaqiang and Song, Fengqi and Cheng, Ran and Chang, Cui-Zu},
  journal = {Nano Lett.},
  volume = {24},
  number = {27},
  pages = {8222--8229},
  year = {2024},
  doi = {10.1021/acs.nanolett.4c01625}
}

@article{PhysRevB.103.024414,
  title = {Magnetic exchange interactions in the van der Waals layered antiferromagnet $\mathrm{Mn}\mathrm{P}{\mathrm{Se}}_{3}$},
  author = {Calder, S. and Haglund, A. V. and Kolesnikov, A. I. and Mandrus, D.},
  journal = {Phys. Rev. B},
  volume = {103},
  issue = {2},
  pages = {024414},
  numpages = {9},
  year = {2021},
  month = {Jan},
  publisher = {American Physical Society},
  doi = {10.1103/PhysRevB.103.024414},
  url = {https://link.aps.org/doi/10.1103/PhysRevB.103.024414}
}

@article{liu_layer-dependent_2022,
	title = {Layer-dependent magnetic phase diagram in {FenGeTe2} (3 ≤ n ≤ 7) ultrathin films},
	volume = {5},
	issn = {2399-3650},
	url = {https://doi.org/10.1038/s42005-022-00921-3},
	doi = {10.1038/s42005-022-00921-3},
	number = {1},
	journal = {Communications Physics},
	author = {Liu, Qinxi and Xing, Jianpei and Jiang, Zhou and Guo, Yu and Jiang, Xue and Qi, Yan and Zhao, Jijun},
	month = jun,
	year = {2022},
	pages = {140},
}

@article{roemer_unraveling_2024,
	title = {Unraveling the electronic structure and magnetic transition evolution across monolayer, bilayer, and multilayer ferromagnetic {Fe3GeTe2}},
	volume = {8},
	issn = {2397-7132},
	url = {https://doi.org/10.1038/s41699-024-00499-0},
	doi = {10.1038/s41699-024-00499-0},
	number = {1},
	journal = {npj 2D Materials and Applications},
	author = {Roemer, R. and Lee, D. H. D. and Smit, S. and Zhang, X. and Godin, S. and Hamza, V. and Jian, T. and Larkin, J. and Shin, H. and Liu, C. and Michiardi, M. and Levy, G. and Zhang, Z. and Green, R. J. and Kim, C. and Muller, D. and Damascelli, A. and Han, M. J. and Zou, K.},
	month = sep,
	year = {2024},
	pages = {63},
}

@article{Yang2020FeCl2Strain,
  title        = {Strain modulated ferromagnetic phase transitions in monolayer {FeCl}\textsubscript{2} through exchange competitions: the first-principle and Monte Carlo simulations},
  author       = {Yang, Ya and Guo, Peiyin and Luo, Yongsong},
  journal      = {Physical Chemistry Chemical Physics},
  year         = {2020},
  volume       = {22},
  pages        = {17291--17298},
  doi          = {10.1039/D0CP01422B},
  publisher    = {Royal Society of Chemistry},
  issn         = {1463-9076},
  url          = {https://doi.org/10.1039/D0CP01422B},
  note         = {Published 03 Jul 2020}
}

\end{document}